%% file: main.tex
\documentclass[11pt,letterpaper]{article}
\usepackage[margin=1in]{geometry}
\usepackage[T1]{fontenc}
\usepackage{lmodern,microtype,amsmath,amssymb,amsthm,mathtools}
\usepackage{booktabs,tabularx,array,enumitem,xcolor,float,algorithm,algpseudocode,titlesec}
\usepackage{accsupp}
\usepackage[hidelinks]{hyperref}
\usepackage[nameinlink,noabbrev]{cleveref}
\makeatletter
\def\theHALG@line{\thealgorithm.\arabic{ALG@line}}
\makeatother
\hypersetup{pdftitle={Uniform Guarded Bootstrapping for Topology-Sensitive Directed Shortest Paths},pdfauthor={Bin Cai}}
\newtheorem{theorem}{Theorem}[section]
\newtheorem{lemma}[theorem]{Lemma}
\newtheorem{proposition}[theorem]{Proposition}
\newtheorem{corollary}[theorem]{Corollary}

\theoremstyle{definition}
\newtheorem{definition}[theorem]{Definition}
\newtheorem{contract}[theorem]{Specification}
\newtheorem{remark}[theorem]{Remark}

\crefname{contract}{specification}{specifications}
\Crefname{contract}{Specification}{Specifications}
\newcommand{\RAM}{\operatorname{RAM}}
\newcommand{\KCA}{K_{\mathrm{CA}}}
\newcommand{\ls}{\log_2^*}
\newcommand{\GEN}{\operatorname{GEN}}

\newcommand{\R}{\mathbb R}

\newcommand{\dist}{\operatorname{dist}}
\newcommand{\CA}{\operatorname{CA}}
\newcommand{\Eval}{\textsc{Eval}}

\newcommand{\Inf}{\infty}
\newcolumntype{Y}{>{\raggedright\arraybackslash}X}
\setlist{itemsep=3pt,topsep=5pt}
\let\small\normalsize
\title{Uniform Guarded Bootstrapping for\\Topology-Sensitive Directed Shortest Paths}
\author{Bin Cai\\Wuhan University of Technology\\Wuhan, China\\\texttt{369028@whut.edu.cn}}
\date{3 November 2026}
\begin{document}
\hypersetup{pageanchor=false}
\begin{titlepage}
\maketitle
\begin{abstract}
On a source star, every leaf distance is an input weight. Thus computing labeled distances need not reveal their order. For a rooted directed topology, let $K$ be the least worst-case sum of comparisons and additions among finite deterministic programs specialized to that topology and correct for every nonnegative weighting.

We give one deterministic program that takes the graph as input. This graph-input-uniform algorithm has total comparison--addition RAM time
\[
 O\!\left(m+n+K\Phi(K)\exp(O(\log_2^*(K+2)))\right),
\]
where $n$ and $m$ count vertices and arcs, and $\Phi$ is a regularized product of successive logarithms. An active core is the small graph remaining after directly recoverable distances are separated. The algorithm chooses its parameters from the core size.

Frozen entry ceilings let each interrupted local call be replayed on a small completion, whose complete-run bound controls the stopped work. The cost coefficient satisfies a linear one-step recurrence, yielding at most exponential growth in the number of generations. This permits a depth chosen from the measured core size.

For $m=O(n)$, the solver gives $O(n\sqrt{\log n}(\log\log n)^{1/4+o(1)})$ time, improving the sparse directed bound of Duan, Mao, Shu, and Yin. A complementary result uses reusable dominator calibration for changing seed vectors on a fixed weighted graph.
\end{abstract}
\thispagestyle{empty}
\end{titlepage}
\hypersetup{pageanchor=true}
\setcounter{page}{1}
\input{sections/introduction}
\input{sections/core_argument}
\input{sections/uniform_bound}
\input{sections/model}

\input{sections/engine}
\input{sections/bootstrap}
\input{sections/ownership}
\input{sections/calculus}
\input{sections/selector}
\input{sections/consequences}
\input{sections/reuse}
\section{Conclusion}\label{sec:conclusion}

Guarded bootstrapping yields the topology-sensitive running-time bound of \cref{thm:main}. Small completions bound interrupted acquisitions, and the main-call budget gives a linear one-step recurrence for their cost coefficients. Choosing the maximal legal depth from the measured core size reduces the second-logarithm exponent in the sparse directed SSSP bound from $1/2$ to $1/4+o(1)$. The quotient-solver interface also supports the reusable seeded-query composition.

\paragraph{Open problems.}
Can the factor $\exp(O(\log_2^*(K+2)))$ be removed while retaining a uniform algorithm? A sharper coefficient recurrence or a more direct representation of nested calls could improve this overhead. Further questions are whether a richer query benchmark can account for the leaf term, how the hierarchy performs with practical thresholds, and which parts of the completion argument extend to dynamic or approximate shortest paths.
\section*{AI Methodology}
ChatGPT was used only to assist with writing and formatting the final manuscript and checking related content. All ChatGPT-assisted material was manually reviewed and verified by the author, who takes responsibility for the final manuscript.
\clearpage
\appendix
\titleformat{\section}{\normalfont\Large\bfseries}{Appendix \thesection}{1em}{}

\input{sections/guarded_base_appendix}
\input{sections/bridge_proofs}
\input{sections/seeded/introduction}
\input{sections/seeded/calibration}
\input{sections/seeded/composition}
\input{sections/seeded/backend}
\input{sections/seeded/realization}
\input{sections/seeded/complexity}
\input{sections/scope}
\begingroup\small
\bibliographystyle{alpha}
\bibliography{references}
\endgroup
\end{document}

%% file: sections/introduction.tex
\section{Introduction}\label{sec:introduction}

The single-source shortest-path problem asks for the distance from a source to every vertex of a weighted graph. These are \emph{labeled distances}: each value is associated with its vertex, but the values need not be sorted. On a directed star with one arc from the source to each leaf, the leaf distances are the input weights themselves. Returning those values requires no numerical operation, whereas ordering the vertices by distance requires sorting arbitrary nonnegative real numbers.

For directed graphs with nonnegative real weights, recent algorithms exploit this distinction to go beyond Dijkstra's worst-case bound. Duan, Mao, Mao, Shu, and Yin broke the sorting barrier with a deterministic $O(m\log^{2/3}n)$ algorithm in the comparison--addition model~\cite{duan2025breaking}. Duan, Mao, Shu, and Yin (DMSY) subsequently obtained
\begin{equation}\label{eq:dmsy-intro}
 O\!\left(m\sqrt{\log n}+\sqrt{mn\log n\log\log n}\right),
\end{equation}
which is $O(n\sqrt{\log n\log\log n})$ when $m=O(n)$~\cite{duan2026faster}. These bounds measure the hardest graphs of a given size. We seek a bound that also reflects the numerical difficulty of each topology.

\paragraph{A numerical benchmark for a fixed topology.}
Fix a rooted directed topology $(G,s)$. Let $K=\KCA(G,s)$ be the least worst-case sum of comparisons and additions among finite deterministic programs specialized to $(G,s)$ and correct for every independent nonnegative weighting. A competitor may share intermediate sums across outputs and compare arbitrary materialized registers. The output is indexed by vertex, and $K$ measures its joint numerical cost.

Our algorithm receives the graph as input and uses the active-core reduction of~\cite{cai2026beyond}. In $O(m+n+K)$ RAM time, this reduction separates directly recoverable distances and constructs a remaining shortest-path instance of size $O(K)$.

\subsection{Main result}

Let $\Phi$ be the regularized iterated-logarithm product in \eqref{eq:phi-intro}.
\begin{theorem}[Uniform maximal-depth bound]\label[theorem]{thm:main}
In the comparison-addition RAM of \cref{sec:model}, the guarded-prefix construction gives one finite deterministic graph-input-uniform exact SSSP program $U_*$ and absolute constants $a,b>0$ such that for every finite directed multigraph, source, and independent nonnegative real weighting,
\begin{equation}\label{eq:main}
 \RAM_{U_*}(G,s,w)
 \le a\left[m+n+K\Phi(K)
                       \exp\!\bigl(b\ls(K+2)\bigr)\right],
 \qquad K=\KCA(G,s).
\end{equation}

\end{theorem}

For $m=O(n)$, the same solver runs in
$O(n\sqrt{\log n}(\log\log n)^{1/4+o(1)})$ time, reducing the second-logarithm exponent in the DMSY bound from $1/2$ to $1/4+o(1)$. The topology-sensitive bound and the complementary seeded-query result are collected below.
\begin{table}[H]
\centering
\caption{Running-time bounds. For seeded queries, $h$ is the number of dominator-tree leaves and $\alpha$ is an inverse-Ackermann bound.}\label{tab:contributions}
\begin{tabularx}{\textwidth}{@{}>{\raggedright\arraybackslash}p{.32\textwidth}Y@{}}\toprule
Problem & Total RAM time\\\midrule
Topology-sensitive SSSP (\cref{thm:main}) & $O(m+n+K\Phi(K)\exp(O(\ls(K+2))))$\\\addlinespace[3pt]
Sparse directed SSSP (\cref{cor:sparse}) & $O(n\sqrt{\log n}(\log\log n)^{1/4+o(1)})$\\\addlinespace[3pt]
Seeded queries after preprocessing (\cref{sd:thm:main}) & $O((m+n)\alpha(m+n+2)+h\log(h+1))$ per query\\\bottomrule
\end{tabularx}
\end{table}

The new step is to replace local growth by a fresh call to the preceding generation. A small completion bounds the work before an external interruption. A main-call budget then yields $C_{d+1}\le B(C_d+1)$ and hence $C_d\le AB^d$. Choosing the largest legal generation gives the factor $\Phi$ up to an exponential in log-star.

\subsection{Relation to previous work}

\paragraph{Distance ordering and universal optimality.}
Haeupler, Hlad\'{\i}k, Rozho\v{n}, Tarjan, and T\v{e}tek prove that Dijkstra's algorithm with a suitable working-set heap is universally optimal for \emph{distance ordering}~\cite{haeupler2024universal}. Their benchmark requires a distance order; ours measures the joint numerical cost of labeled distances. The source-star example separates these tasks. Recent simpler and pointer-machine implementations retain the distance-order task~\cite{hoog2026simple,hoog2026pointer}.

\paragraph{From local search to a uniform hierarchy.}
DMSY combine bounded multi-source recursion with local Dijkstra searches, edge-disjoint forest grouping, and pivot representatives in a partial-sorting data structure~\cite{duan2026faster}. We retain this recursion and grouping structure. Our guarded hierarchy replaces the local searches by fresh calls to the preceding generation. Frozen entry ceilings and small completions control externally interrupted calls; a stronger recursive budget yields a linear one-step recurrence for their cost coefficients. The storage and parameter bounds are uniform in depth, so the generation is chosen from the input size. The guarded base and the generation induction are proved in \cref{gb:appendix,sec:bootstrap}. Zhou develops a Deadline-BMSSP bootstrap~\cite{zhou2026deadline}.

Grishin and Zabokritskiy give a related packed-search algorithm for directed SSSP~\cite{grishin2026bounds}. Their paper appeared on arXiv on 19 November 2026, after the public posting of this manuscript as STOC 2027 submission \#1592 on 2 November 2026 (11:59 pm AoE). They prepare short walks and use bit-mask operations to accelerate inner searches, obtaining an $O(n\sqrt{\log n}(\log\log n)^{1/4})$ bound on sparse graphs and additional bounds at intermediate densities. Both works analyze searches that can stop before completion. Their Proposition 7.5 compares a stopped packed invocation with a completed run on an induced retained subgraph; our small-completion lemma bounds an interrupted lower-generation call using frozen entry ceilings. The papers therefore overlap in the problem and in the accounting question for stopped searches, while using different constructions and overall acceleration mechanisms.

A preliminary preprint~\cite{cai2026beyond} developed the resource framework, numerical optimal synthesis, and active-core reduction. The present work builds on that foundation by introducing the guarded hierarchy, which yields the topology-sensitive bound in \cref{thm:main}.

\paragraph{Structural decompositions.}
Tarjan's path-expression methods use dominator structure and reusable evaluation~\cite{tarjan1981}. Acyclic-connected trees provide a different structural parameter and recursive shortest-path transformation~\cite{stefansson2026}. Our seeded-query extension uses the number of dominator-tree leaves and reuses numerical calibration across queries. Kadria and Roditty give an improved bound for undirected graphs~\cite{kadria2026}; the main result here concerns directed labeled distances.

%% file: sections/core_argument.tex
\section{Guarded bootstrapping}\label{sec:overview}

\subsection{Local growth and fresh calls}\label{sec:overview-growth}\label{sec:acquisition}

We use DMSY's frontier recursion and forest grouping~\cite{duan2026faster}. The frontier contains complete vertices witnessing the shortest paths to the remaining targets. Local growth organizes it into groups with connected witness trees, each represented by a pivot. A growth attempt starts at one root and has three possible outcomes: it discovers a prescribed number of vertices, first meets the existing forest, or exhausts its admissible region. The first two outcomes enlarge the witness forest; exhaustion certifies the closure needed to remove a failed root. We call this the \emph{Witness-or-Seal} (WOS) step. Guarded bootstrapping preserves these outcomes while replacing the local Dijkstra search by a recursive solver.

A \emph{generation} records how many times we have replaced this local growth routine by a lower-generation solver. It differs from the level of the main recursion: within one generation, main calls partition the target into smaller distance intervals; a new generation is entered only to carry out local growth.

The lower solver explores a guarded problem. Every vertex has a fixed ceiling, and every prefix of an admissible root path must satisfy its endpoint ceiling. These ceilings combine the enclosing constraints with the labels present when the call begins. Starting from a single root gives a well-defined local problem even though its discoveries are published to enclosing calls as they occur.

After this substitution, a growth stop can interrupt a nested recursive computation. Suppose the lower call has accepted $a$ vertices at that point. Its complete-run theorem bounds cost in terms of the eventual output, which may be much larger than $a$ on the original graph. The interruption may also occur inside an amortized data-structure operation. We therefore need a bound on the actual execution prefix in terms of $a$, with the parameters chosen at entry still fixed. The completion argument below relates this prefix to a complete run with a small output and the same entry parameters.

Discoveries are counted separately in the fresh engine and in the enclosing growth attempt. Let $S$ be the lower engine's accepted set and $A$ the enclosing discovery union. Every lower discovery is published through each live receiver before the next graph operation. At an enclosing quota $q$,
\begin{equation}\label{eq:union}
 S\subseteq A,\qquad |S|\le |A|\le q.
\end{equation}
This separation permits equal-valued rediscovery in a fresh lower engine while counting that vertex only once in an enclosing union.

\subsection{Bounding an interrupted call}\label{sec:small-completion}\label{sec:completion}

Assume inductively that generation $d$ satisfies the complete-run bound \eqref{eq:raw}, the guarded-key convention \ref{b:keys}, the locality condition \ref{b:local}, and the interruption rule \ref{b:stops}. Write $F=f_d$ and $g=f_{d-1}$ for the generation factors defined in \cref{sec:overview-factors}. At outer width $t$, the acquisition uses a quota $q=\Theta(t/F(t))$, lower width $s=\Theta(F(q))$, capacity $Q^*=8(q+1)$, and level $\ell_{\rm pkt}\le c_0+c_1L_1(q)/s$. The widths are legal and satisfy $F(q)=\Theta(F(t))$ and $qF(q)=\Theta(t)$; the integer choices appear in \cref{sec:calculus,sec:engine}.

Let $S$ be the accepted set when the lower call stops and $a=|S|\le q$. Its small completion retains every incident arc and endpoint, together with complete queried lists, original IDs, adjacency order, and the last partially scanned list. At in-degree and out-degree at most three,
\begin{equation}\label{eq:incident}
 |S|+|\operatorname{heads}(E^+(S))|
     +|\operatorname{tails}(E^-(S))|\le7|S|.
\end{equation}

Keep the entry parameters fixed. Preserve physical IDs, registers, list order, offsets, relevant frozen ceilings, and the chosen packet. Every queried list belongs to an accepted vertex by \ref{b:local}; transitive publication ensures this also for deeper engines. Every list entry and numerical comparison input therefore agrees. Every branch uses the held entry packet and the retained queried data. Induction on operations identifies the two stopped executions, with a bijection between their freshly allocated records.

Only after matching this prefix, replace the raw boundary's external receiver by a bounded no-stop receiver, or suppress all its external stop actions. This includes higher ancestors' foreign quotas. Keep every stop and partial return native to an engine inside the raw boundary. Frozen ceilings remain unchanged: relevant entry values were captured before upward writes. The continued call therefore has the same guarded problem. The root output threshold exceeds $Q^*>7a$, forcing a full return on the completed graph under its arbitrary-level theorem.

The whole completion costs at most an absolute multiple of
\begin{equation}\label{eq:completioncost}
 C_d(1+a)[s+\ell_{\rm pkt}g(s)+(\log_2s)^2].
\end{equation}
All raw internal prefix work and its unit boundary requests are nonnegative work inside that completion under the stated cost convention. The enclosing receiver-link costs are paid by their respective wrappers in the linear composition below. Abort propagation and export cleanup are separately linear in earlier allocations and $a$ by \ref{b:stops}; the same majorant pays them. Bounding the prefix by a complete execution also accounts for unfinished data-structure operations.

Using the level estimate $\ell_{\rm pkt}\le c_0+c_1L_1(q)/s$, \cref{lem:packet} gives
\begin{equation}\label{eq:acquisitionW}
 \mathcal{T}_{\rm acq}(q,a)\le W_{d+1}(1+a)F(q),\qquad
 W_{d+1}\le a_0+a_1C_d
\end{equation}
for absolute constants. The legality estimates in \cref{sec:calculus} give $F(q)\ge T_d\ge d+1$, which pays for the $O(d+1)$ packet evaluation. The new wrapper adds one receiver/capture link with constant work per lower event; each enclosing wrapper pays for its own link. Parameter evaluation contributes to $a_0$, and the added link contributes to $a_1$.

\subsection{The main-call budget}\label{sec:coefficient}

Fix a generation-$(d+1)$ main call $X$, with width $t$, level $\ell_X$, and source set $S_X$. Write $\pi(v)$ for the fixed optimum key of vertex $v$ in this engine's guarded problem. Put $F=f_d$ and $W=W_{d+1}$. Let $U=U_X$ be its output, $Q_X$ its failed growth roots, $X_i$ its same-generation children, $U_i$ their outputs, $B_i$ their input bounds, and $W'$ its final complete remainder. The main-call subtree consists of calls in this engine; lower-generation growth is charged by \eqref{eq:acquisitionW}. Write $B_X$ and $\beta_X$ for the input and return bounds of $X$.

For a vertex set $Z$, let $N_I(Z)$ contain the primitive arcs with complete tails in $Z$ whose extended optimum keys lie in $I$, including ceiling-rejected proposals. Let $E(Z)$ contain the arcs with both endpoints in $Z$, and put $E_{\rm cut}=E(U)\setminus\bigcup_iE(U_i)$. We prove the bound
\begin{equation}\label{eq:strongbudget}
\begin{split}
 P_C(X)={}&Cu_X[\ell_XF(t)+(\log_2t)^2]\\
 &+Ct\left[|N_{[\mu_X,B_X)}(U_X)|+|E(U_X)|
                   +\sum_{Y\text{ in subtree}(X)}|Q_Y|\right],
\end{split}
\end{equation}
where $\mu_X=\min(\{B_X\}\cup\pi(S_X))$ and $u_X=\max\{1,|U_X|\}$.

\paragraph{The budget left by the children.}
A legal call has a nonempty source set. A full return contains that set, and a partial return exceeds its positive output threshold (\cref{gb:inputs,gb:prefix}). Thus $u_X=|U_X|$ and $u_i=|U_i|$ for the calls in this induction. At level zero, BASE costs $O(u_X\log t)$, paid by the squared-logarithm term. For $\ell_X\ge1$, the outputs partition as $U=\bigsqcup_iU_i\sqcup W'$, and every child has level $\ell_X-1$. Writing $u=|U|$ gives
\begin{align*}
 &u[\ell_XF(t)+(\log_2t)^2]
   -\sum_i u_i[(\ell_X-1)F(t)+(\log_2t)^2]\\
 &\qquad=uF(t)+|W'|[(\ell_X-1)F(t)+(\log_2t)^2]
 \ge uF(t).
\end{align*}

Set
\[
 R_4=\sum_i|N_{[B_i,B_X)}(U_i)|,
 \qquad R_7=|N_{[\beta_X,B_X)}(W')|.
\]
The progress invariant gives $\mu_i\ge\mu_X$ and $\beta_X\ge\mu_X$ (\cref{gb:target,gb:prefix}). For a tail in $U_i$, its child interval $[\mu_i,B_i)$ and parent interval $[B_i,B_X)$ are disjoint and contained in $[\mu_X,B_X)$. Different children have disjoint tail sets; the final interval $[\beta_X,B_X)$ uses tails in $W'$. Consequently,
\[
 |N_{[\mu_X,B_X)}(U)|
 -\sum_i|N_{[\mu_i,B_i)}(U_i)|\ge R_4+R_7.
\]
The internal-edge sets satisfy $E(U)=\bigsqcup_iE(U_i)\sqcup E_{\rm cut}$. The failed-root subtree sum partitions into the child sums and $|Q_X|$. Subtracting the child budgets in \eqref{eq:strongbudget} therefore leaves
\begin{equation}\label{eq:budget-slack}
 P_C(X)-\sum_iP_C(X_i)
 \ge CuF(t)+Ct\bigl(R_4+R_7+|E_{\rm cut}|+|Q_X|\bigr).
\end{equation}

\begin{table}[H]
\centering\small
\renewcommand{\arraystretch}{1.18}
\caption{Local costs and their sources in the budget difference \eqref{eq:budget-slack}. The constant $c$ is absolute.}\label{tab:accounts}
\begin{tabularx}{\textwidth}{@{}>{\raggedright\arraybackslash}p{.29\textwidth}>{\centering\arraybackslash}p{.32\textwidth}Y@{}}\toprule
Account & Cost bound & Charged by\\\midrule
A1: growth, partition, initial groups and pivots & $c(W+1)[uF(t)+t|Q_X|]$ & Level decrease $CuF(t)$ and local failed roots $Ct|Q_X|$\\
A2--A3, A6: queue work, scans, cleanup, residual sources & $cu$ & Level decrease $CuF(t)$\\
A4: external insertions & $ctR_4$ & Parent boundary intervals $CtR_4$\\
A5: full replacements & $cuF(t)+ct|E_{\rm cut}|$ & Level decrease and cut edges\\
A5: partial replacements & $cu$ & Level decrease $CuF(t)$\\
A7: final boundary insertions & $ctR_7$ & Remaining boundary intervals $CtR_7$\\\bottomrule
\end{tabularx}
\end{table}

\paragraph{Growth and failed roots (A1).}
Each acquisition counts its root, so $1+a\le2a$. Successful acquisitions have total accepted mass $O(u)$ by the connected-growth bound. Their cost is $O(WuF(q))=O(WuF(t))$. Each failed acquisition has accepted mass at most $q$, so \eqref{eq:acquisitionW} charges it $O(WqF(q))=O(Wt)$. Charging each failed root separately gives $O(Wt|Q_X|)$ even when failed closures overlap. The $O(u/q)$ initial groups require $O(tu/q)=O(uF(t))$ pivot work. These terms use the level decrease and the local failed-root part of \eqref{eq:budget-slack}.

\paragraph{Boundary attempts (A4 and A7).}
The scan after child $X_i$ returns examines each outgoing primitive arc of $U_i$ once. An expensive external Insert is made when its candidate belongs to $[B_i,B_X)$; each such attempt has amortized cost $O(t)$ (\cref{alg:raw,gb:cost}). This gives $O(tR_4)$. Attempts reaching the same head are paid separately by their distinct primitive arcs, including parallel arcs. For a fixed primitive arc, the half-open child and parent intervals allocate its external Insert charge to a single node of this engine's recursion tree. The ordinary relaxation tests performed at different ancestors belong to the linear scan account A3. The final scan uses the disjoint tail set $W'$ and inserts at candidates in $[\beta_X,B_X)$, giving $O(tR_7)$. The interval subtraction above supplies both amounts, including unsuccessful duplicate-key insertion tests.

\paragraph{Replacement pivots (A5).}
For a full call, let $\mathcal F_j$ be the edge-disjoint witness tree of group $j$ contained in $U$, and let $g_j$ count child iterations that remove its designated pivot and require a replacement. The removed pivots belong to $g_j$ different child outputs. Deleting the edges of $\mathcal F_j\cap E_{\rm cut}$ puts these pivots in distinct components, so
\[
 g_j-1\le |\mathcal F_j\cap E_{\rm cut}|,
 \qquad \sum_jg_j\le p+|E_{\rm cut}|,
\]
where $p=O(u/q)$ is the initial number of groups. A replacement scans $O(q)$ group vertices and inserts one pivot at amortized cost $O(t)$. Since $q=O(t)$, all replacements cost $O(t\sum_jg_j)=O(uF(t)+t|E_{\rm cut}|)$. The identity for $E(U)$ makes the cut-edge charge telescope: an edge enters $E_{\rm cut}$ at the first node whose output contains both endpoints and none of whose child outputs contains both endpoints. It receives at most one cut-edge charge in this main-call tree. For a partial call, each removed source is used once and $t|S_X|\le u$, giving the linear bound in the table.

\paragraph{Closing the two inductions.}
The remaining queue and scan costs are $O(u)$ by \cref{gb:cost}, and $F(t)\ge1$. Choose $C\ge c_*(W+1)$ for a sufficiently large absolute $c_*$. The difference \eqref{eq:budget-slack} pays every row of the table, while induction on the main-recursion level pays each child by $P_C(X_i)$. Therefore the total is at most $P_C(X)$. The same $C$ appears at every node of this main recursion: $W$ occurs in the local acquisition cost A1, and the child budgets are added with coefficient one.

Finally, \eqref{eq:Qbound} and bounded degree give $\sum_Y|Q_Y|\le u(\Delta+1+\ell_X/t)$ and $|N_I(U)|,|E(U)|\le\Delta u$. Substitution in \eqref{eq:strongbudget} yields
\[
 \mathcal{T}_{d+1}(X;t,\ell)\le C_{d+1}u_X
          [t+\ell F(t)+(\log_2t)^2],\qquad
 C_{d+1}\le b_0+b_1W_{d+1}\le B(C_d+1).
\]
This is a linear one-step recurrence. Its solution satisfies $C_d\le AB^d$ for absolute $A,B>1$.

%% file: sections/uniform_bound.tex
\section{Uniform depth and topology sensitivity}\label{sec:main}

\subsection{The logarithmic hierarchy}\label{sec:overview-factors}

For $x\ge0$ and integers $d\ge0$, define the regularized iterated logarithms and their products by
\begin{equation}\label{eq:logs-intro}
 L_1(x)=\log_2(2+x),\qquad L_{i+1}(x)=\log_2(4+L_i(x)),
\end{equation}
\begin{equation}\label{eq:fd-intro}
 f_d(x)=\left(\prod_{i=1}^dL_i(x)^{2^{-i}}\right)
       (L_{d+1}(x)L_{d+2}(x))^{2^{-(d+1)}},
\end{equation}
and
\begin{equation}\label{eq:phi-intro}
 \Phi(x)=\prod_{i\ge1}L_i(x)^{2^{-i}}.
\end{equation}
Multiplying the argument by a fixed positive constant changes every regularized iterate by at most a constant factor, uniformly in its index. Since the exponents have total mass one, the same stability holds for $f_d$ uniformly in $d$, and for $\Phi$. Rounding and parameter slack therefore cost a constant factor independent of depth.

The first factors are $f_0=(L_1L_2)^{1/2}$ and $f_1=L_1^{1/2}(L_2L_3)^{1/4}$. A further generation splits the final exponent between two successive logarithms. The exact identity
\[
 f_{d+1}(x)^2=L_1(x)f_d(L_1(x)+2)
\]
expresses the balance between the new main width and the lower-generation work. In the root bound, the width-dependent terms are
\[
 t+\frac{L_1(x)}{t}f_{d-1}(t)+(\log_2t)^2.
\]
The first term grows with the width; the second combines the number of main-recursion levels with the lower-generation rate. Choosing $t=\Theta(f_d(x))$ balances these two terms and absorbs the squared logarithm, giving $O(M+N+MAB^df_d(x))$ ideal events on a normalized graph with $M$ arcs and $N$ vertices. Here $x=M+N+2$. \Cref{sec:calculus} proves these estimates. For each fixed $x$, the finite factors $f_d(x)$ converge to $\Phi(x)$ as $d\to\infty$.

\subsection{The largest legal generation}\label{sec:overview-depth}

We use $\log_2^*x$ for the least number of binary logarithms needed to reach a value at most one, for $x>1$.

Let $T_4$ be the fixed base threshold in \cref{con:engine}. Define
\begin{equation}\label{eq:thresholds-intro}
 T_{d+1}=2^{64(T_d+1)^2},\qquad X_d=T_{d+1}.
\end{equation}
For $x\ge X_4$, define
\begin{equation}\label{eq:depth}
 D_*(x)=\max\{d\ge4:T_{d+1}\le x\}.
\end{equation}
The thresholds increase strictly and without bound, so this maximum exists. Put $A_0=256(T_4+1)^2$.

\begin{theorem}[Uniform near-envelope truncation]\label[theorem]{thm:truncation}
For $x\ge X_4$ and $d=D_*(x)$,
\begin{equation}\label{eq:truncation}
 1\le\frac{f_d(x)}{\Phi(x)}
       \le A_0^{2^{-d}}\le A_0.
\end{equation}
In particular the ratio tends to one as $x\to\infty$.
\end{theorem}

\begin{lemma}[Threshold height]\label[lemma]{lem:height}
For sufficiently large $x$,
\begin{equation}\label{eq:height}
 D_*(x)=\ls x+O(1).
\end{equation}
All constants depend only on $T_4$.
\end{lemma}

\Cref{sec:selector} gives the integer selector and the proofs of these two bounds. In particular $f_{D_*(x)}(x)/\Phi(x)\to1$, while $B^{D_*(x)}=\exp(O(\log_2^*x))$.

\subsection{From the active core to the global bound}
A vertex $v$ dominates $u$ when every source-to-$u$ path contains $v$. Delete each arc $(u,v)$ with $v$ dominating $u$. Let $k$ be the number of remaining arcs whose tail is not the source and let $p_v$ be the multiplicity of the remaining source-to-$v$ class. Put $\sigma=\sum_{v:p_v>0}(p_v-1)$. The active core $H$ retains the endpoints of the $k$ arcs and their source-class minima. The reduction of~\cite{cai2026beyond} has the following bound; \cref{sec:active-core} gives the construction and \cref{br:active-charge} proves its numerical charge.
\begin{contract}[Active-core reduction with recovery]\label[contract]{con:core}
On the reachable topology, the reduction and full-label lifting cost $O(m+n+K)$ RAM. If $k=0$, all outputs are returned without a core solve. Otherwise,
\begin{equation}\label{eq:core-size}
 |E(H)|\le3k,\quad |V(H)|\le2k+1,\quad
 \sigma+\frac{k}{2}\le K.
\end{equation}
Linear degree normalization creates a graph of maximum in-degree and out-degree at most three, with $M$ arcs and $N$ vertices. Its cost includes zero arcs, source registers, decoding, and all numerical and RAM operations. The constructed graph satisfies
\begin{equation}\label{eq:x}
 x=M+N+2\le c_{\rm core}(K+1),\qquad M=O(K)
\end{equation}
for absolute constants when a nonempty core is solved.
\end{contract}

\begin{proof}[Proof of \cref{thm:main}]
By \cref{prop:base-realized,prop:reconstruct}, the guarded base and its fresh-prefix induction satisfy \cref{con:engine}. Identify the reachable subgraph, apply the paid active-core reduction of \cref{con:core}, and handle the empty-core case directly. On a nonempty degree-normalized core measure $x=M+N+2$. Use \textsc{MaxLegalDepth}$(x)$; invoke the fixed small-core routine if requested, and otherwise run $\GEN(D_*(x),\cdot)$ with its main width.

Let $d=D_*(x)$ in the latter case. The selector and \cref{prop:reconstruct} give a legal execution at this depth.

The raw coefficient, the core-root account, and \cref{lem:balance} give
\[
 E_{\rm core}=O\bigl(M+N+MAB^df_d(x)\bigr).
\]
Here $E_{\rm core}$ counts ideal events. \Cref{prop:reconstruct} gives the RAM realization; since $N\le M+1$ on a nonempty reachable core, its total cost has the same form for suitable absolute $A,B$. By \cref{thm:truncation,lem:height}, for absolute constants $a_1,b_1$,
\begin{equation}\label{eq:Hbound}
 H_*(x):=AB^{D_*(x)}f_{D_*(x)}(x)
 \le a_1\Phi(x)\exp\!\bigl(b_1\ls(x+2)\bigr).
\end{equation}
Extend $H_*$ by a fixed bounded majorant on the small-core region; it also covers the selector's metadata work.

Define the monotone envelope
\begin{equation}\label{eq:hull}
 g_*(K)=1+\max_{1\le z\le\lfloor c_{\rm core}(K+1)\rfloor}H_*(z).
\end{equation}
The maximum is over integer scales. Using $M=O(K)$ and the wrapper gives
$O(m+n+K g_*(K))$. The product $\Phi$ is monotone, and \cref{lem:stability} gives uniform stability under the fixed multiplier $c_{\rm core}$. Also
$\ls(c_{\rm core}(K+1)+2)=\ls(K+2)+O(1)$. Thus \eqref{eq:Hbound} uniformly over the range in \eqref{eq:hull} proves
\[
 g_*(K)\le a_2\Phi(K)\exp\!\bigl(b_2\ls(K+2)\bigr).
\]
For the empty core, $k=0$ and direct recovery costs $O(m+n+K)$. Together with the preceding bound for nonempty cores, this proves \eqref{eq:main}.
\end{proof}

%% file: sections/model.tex
\section{Numerical model and core construction}\label{sec:model}

\subsection{The numerical task}\label{sec:numerical-task}

Let $G=(V,E)$ be a finite directed multigraph with source $s$, $n=|V|$, and $m=|E|$. Each primitive arc has an independent input weight $w_e\in\R_{\ge0}$. We allow parallel arcs, loops, zero weights, ties, and zero-weight cycles. The required output is the labeled vector $(d_w(s,v))_{v\in V}$, with a numerical register for each finite value and a discrete infinity tag for an unreachable vertex. A topological reachability scan costs $O(m+n)$, after which numerical work can be confined to the reachable subgraph. Nonnegativity ensures that removing closed subwalks never increases path length.

A topology-specialized program has finite code, finitely many fixed real literals, and finitely encoded discrete information about its fixed topology. The model distinguishes three kinds of cost.

\paragraph{Charged numerical instructions.} The numerical instruction set consists of binary addition and three-outcome comparison of materialized finite registers. Each instruction costs one, and literal zero is available.

\paragraph{Free benchmark operations.} Copying and selecting a held register are free when driven by topology, discrete state, and charged comparison outcomes. The uniform solver pays for these operations in RAM time.

\paragraph{Metadata and total time.} A uniform solver is one finite program receiving the graph as input. It also pays for graph access, allocation, indexing, parameter calculations, cleanup, and explicit output. Numerical registers have unit cost. Discrete indices use $O(\log(m+n+2))$-bit words; core scale metadata use $O(\log(x+2))$ bits in addition to input pointers. Integer addition, comparison, multiplication, division, shifts, and bit length take unit time on these words, with fixed headroom. Integer operations act on discrete metadata; real-weight registers support the numerical instructions above.

A competitor may add arbitrary held values, compare mixed sums, reuse variables and registers, share computations across outputs, and branch adaptively on equality. Correctness is required on the entire independent nonnegative orthant. If $C_A(w)$ and $P_A(w)$ count its comparisons and additions, respectively, define
\begin{equation}\label{eq:K}
 \KCA(G,s)=\min_A\sup_{w\in\R_{\ge0}^{E}}
                    \bigl(C_A(w)+P_A(w)\bigr).
\end{equation}
The minimum is over the topology-specialized programs just described and is attained by the integer-valued argument in Lemma~\ref{br:attainment}.

\paragraph{Engine cost.}\label{sec:parameters}
A raw cost $\mathcal T$ counts internal numerical and discrete events, with unit owner-store access and unit requests to the immediate receiver. Each enclosing call pays its receiver/capture link. \Cref{lem:storage} realizes this event count in RAM time. All constants in the hierarchy bounds are absolute and uniform in generation and depth.

\subsection{Active-core reduction}\label{sec:active-core}

The reduction in \cref{con:core} is constructed as follows. 

A vertex $v$ dominates $u$ if every directed $s$--$u$ path contains $v$. A primitive $(u,v)$ is inactive if $v$ dominates $u$. Such a primitive can only close a nonnegative subwalk on a rooted path. Deleting all inactive primitives preserves all distances and the unrestricted numerical benchmark: a solver can ignore inactive inputs, and a solver for the larger topology can be simulated on the smaller one by supplying literal zero in the deleted input locations. The simulation retains the solver's arbitrary register operations.

Among the remaining source primitives, let $p_v$ be the multiplicity of the class $s\to v$, and set
\[
 \sigma=\sum_{v:p_v>0}(p_v-1).
\]
Computing each class minimum uses exactly $p_v-1$ comparisons and no additions. Let $k$ count the active primitives whose tail is not $s$, and let $R$ be their endpoint set. The canonical active core $H$ has source $q$, vertex set $\{q\}\cup R$, the $k$ internal primitives, and one virtual primitive $q\to v$ carrying the already held source-class minimum for each applicable $v\in R$. A vertex outside $R$ is recovered directly from its source-class register. Every virtual weight and recovered output is therefore a materialized numerical value.

The size bounds before normalization follow directly from $|R|\le2k$ and at most one virtual source arc per retained endpoint. Proposition~\ref{br:active-charge} reconstructs the numerical lower charge in \eqref{eq:core-size} on a single execution, using the low-high-order theorem~\cite{georgiadis2012lowhigh}. The structural computation and degree normalization implement the reduction from~\cite{cai2026beyond}, as described below. The precise benchmark lifting also gives
\[
 \KCA(H,q)\le\KCA(G,s)\le\sigma+\KCA(H,q).
\]

The structural wrapper implements the reduction in linear graph-processing time. Compute
the dominator tree using a linear-time RAM algorithm~\cite{georgiadis2012lowhigh},
assign its DFS intervals, and test whether each arc head is an ancestor of
its tail. A linear scan then forms the source-class minima and the endpoint
set $R$. To normalize degrees, replace the incoming incidences of each core
vertex by a zero-weight chain directed toward its central representative,
and its outgoing incidences by a zero-weight chain directed away from that
representative. Each original arc joins its designated outgoing and incoming
ports. All port degrees are bounded by three; there are $O(|E(H)|+|V(H)|)$
ports and zero arcs. Every original walk lifts, and contracting zero chains
maps any normalized walk to an original walk of the same length. Distances
at central representatives are therefore exactly the required core labels.
The initial source is its central representative. Creating ports, assigning
literal zeros, and copying recovered labels take linear time. Source-class
minima cost $\sigma$ comparisons. The numerical core-size lower charge pays
for $\sigma+k$, while the input scan and all discarded labels cost $O(m+n)$.

\input{sections/notation}

%% file: sections/notation.tex
\begin{table}[H]
\centering\fontsize{10}{12}\selectfont
\renewcommand{\arraystretch}{1.12}
\caption{Symbols and terms used in the main argument.}\label{tab:notation}
\begin{tabularx}{\textwidth}{@{}>{\raggedright\arraybackslash}p{.20\textwidth}Y>{\raggedright\arraybackslash}p{.19\textwidth}@{}}\toprule
Symbol or term & Meaning & Definition\\\midrule
$K=\KCA(G,s)$ & Topology-specific optimum for the worst-case joint number of comparisons and additions & \Cref{sec:introduction}; \eqref{eq:K}\\\addlinespace[3pt]
$\Phi(x)$ & Product of regularized iterated logarithms & \eqref{eq:phi-intro}\\\addlinespace[3pt]
$f_d(x)$ & Finite logarithmic factor at generation $d$ & \eqref{eq:fd-intro}\\\addlinespace[3pt]
$\ls x$ & Least number of binary logarithms needed to reach a value at most one, for $x>1$ & This table; \cref{sec:overview-depth}\\\addlinespace[3pt]
$T_d$, $X_d$ & Minimum legal width at generation $d$, and the corresponding core-size threshold $X_d=T_{d+1}$ & \eqref{eq:thresholds-intro}\\\addlinespace[3pt]
WOS & Witness-or-Seal growth: quota or forest contact gives a witness tree; exhaustion gives closure & \Cref{sec:overview-growth}\\\addlinespace[3pt]
Fresh-prefix & A fresh single-root lower call, stopped by the enclosing growth attempt & \Cref{sec:acquisition}\\\addlinespace[3pt]
$d$, $\ell$ & Generation of the solver and level of the main recursion within that generation & \Cref{sec:overview-growth}\\\bottomrule
\end{tabularx}
\end{table}

%% file: sections/engine.tex
\section{Detailed construction of the guarded hierarchy}\label{sec:engine}\label{sec:bootstrap}

An engine stores one generation's state. Its main-recursion calls share that state and return prefixes of the guarded target defined in \cref{gb:inputs}. Each growth attempt opens a fresh engine for the preceding generation. The finite base has the entry and structural properties in \cref{con:base}.

%% file: sections/bootstrap.tex
\subsection{The fixed base and structural invariants}

We work with graphs of degree at most $\Delta=3$. The base has the frontier, locality, and cost properties stated in \cref{con:base}.

\begin{proposition}[Finite guarded basis]\label[proposition]{prop:base-realized}
The finite guarded algorithm in Appendix~\ref{gb:appendix} realizes \cref{con:base}. Its fresh-root top calls satisfy the entry conditions, and its recursively generated calls preserve them.
\end{proposition}
\begin{proof}
A fresh root is complete, so its singleton source set satisfies SEAL. The admissible labels and tight-path order in \cref{gb:inputs} give \ref{b:keys}. \Cref{gb:target,gb:prefix} prove source-minimum equality, ready-frontier coverage, exact returned prefixes and preservation of SEAL for every child. This includes vertices that were already complete at entry.

\Cref{gb:a0} bounds the ordinary guarded call. The stopped-prefix construction in \cref{gb:finite} then gives the finite chain
$A_0\to\mathrm{GP}_0\to A_1\to\cdots\to\mathrm{GP}_3\to A_4$.
\Cref{gb:a4} is precisely the generation-four bound in \ref{b:base}, on legal word-range levels.

Locality follows from the scan order: ordinary acquisition scans only extracted, locally accepted tails, and main/BASE scans use complete output tails. Relaxation publishes acceptance before the next adjacency or queue action. Fresh-prefix substitution preserves this order recursively. Allocation is lazy, and parameters depend on the held packet rather than unread graph lists. Entry-ceiling capture precedes the first outward write at every enclosing owner, as specified in \cref{gb:finite}. Thus all list queries and visible numerical operands occur within the retained incident-prefix data, establishing \ref{b:local}.

The stop/export sequence in \cref{gb:finite} gives \ref{b:stops}: the root keeps its enclosing predecessor, values are copied before retirement, owner-targeted foreign stops propagate, and cleanup visits only touched records and frames. Connected growth preserves the forest partition in \cref{gb:algorithm}; full-failure closure preserves the target and frontier, and \cref{gb:origins} bounds failed roots. This proves \ref{b:structure}. Fresh-prefix acquisition preserves the connected growth and closure used in these arguments. The child capacity follows from $3q+1\le t$.

The indexed BASE and block representation in \cref{gb:cost} provide separated Merge and direct deletion handles. Their nonnegative normalization potential and the budget \eqref{gb:strong} bound duplicate tests, repeated appearances, record creation, and cleanup, giving \ref{b:physical}. Logical owner lookups are converted to physical searches by \cref{lem:storage}.
\end{proof}

\subsection{Fresh-prefix acquisition}

Assume a legal raw generation $d$ already has \eqref{eq:raw} and \ref{b:keys}, \ref{b:local}, \ref{b:stops}. Write $F=f_d$, $g=f_{d-1}$. An outer generation-$d+1$ width $t$ chooses
\[
 q=\left\lfloor t/(4096\widehat f_d(t))\right\rfloor,\quad
 Q^*=8(q+1),\quad s=\widehat f_d(q),\quad
 \ell_{\rm pkt}=\max\{1,\lceil\log_2 Q^*/s\rceil\}.
\]
The logarithm in this integer parameter can be replaced by an upward bit-length value. This changes only absolute level constants.

\begin{algorithm}[H]
\caption{\textsc{FreshPrefix}$(r,q,d)$ in an outer growth attempt}\label{alg:freshprefix}
\begin{algorithmic}[1]
\State Open a fresh lower engine, with only root $r$ finite at the held outer offset. Lazily freeze entry incumbents intersected with old ceilings, capturing each value before any relevant upward write.
\State Count the local root as accepted. Preserve its existing predecessor in the enclosing context; do not export a null lower-root predecessor over that edge.
\State Run raw $\GEN(d,\cdot)$ with packet $(s,\ell_{\rm pkt},Q^*)$. At each publication, copy label and nonroot predecessor to the immediate receiver, record that receiver's own distinct discovery, invoke its outer receiver, then test this attempt's forest contact and quota.
\State At first forest contact return MERGE; at $q$ distinct local acceptances return CAP. Forward a foreign stop to its owner. Only genuine full normal lower-root exhaustion returns FAILURE.
\State Export the accepted values and predecessor arcs, then reclaim touched state.
\end{algorithmic}
\end{algorithm}

Current predecessor arcs form a rooted arborescence: a relaxation installs a strictly smaller parent key, a later parent decrease preserves this inequality, and a later head decrease installs a new parent. A nonnegative cycle cannot improve the local root because deleting it lowers length or hops. The root-initialization exception in line 2 is necessary when the growth root already has an enclosing predecessor. First contact appends only new vertices and arcs to one old tree, costing $O(a)$ for local accepted mass $a$; no old tree is traversed.

Target membership follows from the SEAL invariant in \ref{b:keys}. An accepted candidate outside the target below the bound would enter a vertex already complete by SEAL. Its key must equal that optimum and have an optimal prefix through the root, placing it in the target after all. For FAILURE, exact local closure and frozen path upper bounds preserve every needed tight suffix from a complete root. These are exactly the structural premises of grouping and frontier replacement in \ref{b:structure}. CAP/MERGE supplies connected growth; full FAILURE supplies the tight-path closure.

\begin{proposition}[Uniform guarded engine]\label[proposition]{prop:reconstruct}
The fresh-prefix construction satisfies \cref{con:engine} on every legal packet, with the owner-store representation and absolute constants.
\end{proposition}
\begin{proof}
Induct on generation using \eqref{eq:acquisitionW} and the stronger main budget. Legality and root balance follow from \cref{lem:evaluator,lem:balance}; same-generation calls retain width.

For a reachable normalized graph of measured size $x=M+N+2\ge X_d$, choose
\[
 t=\widehat f_d(x),\qquad
 \ell_{\rm main}=\max\{1,\lceil\operatorname{bitlen}(x)/t\rceil\}.
\]
Start the raw call from the graph source with offset zero, infinite ceilings and infinite bound. The singleton source satisfies SEAL. Its output threshold obeys
$t^3 2^{\ell_{\rm main}t}\ge x>N$.
The partial-output rule of \cref{gb:prefix} therefore forces a full return, with $u_X=N$. Moreover,
$\ell_{\rm main}\le1+L_1(x)/t$ up to a fixed rounding constant. Since $N\le M+1$ and $f_{d-1}(t)\le t$ on legal scales, the raw bound gives exactly the root account in \ref{e:root}, including the additive graph setup.

Fresh generation descent opens at most one engine at each generation, and main frame stacks remain within those payloads, so $L=O(D+1)$ including the finite base. Applying \cref{lem:storage} to the ideal-event bound adds an $O(D+1)$ factor, absorbed in a larger exponential. On a nonempty reachable normalized graph, $N\le M+1$, so its multiplication of the additive linear term is also absorbed in $MA'(B')^D\,f_D(x)$. At the maximal legal depth, $B^D=\exp(O(\ls x))\le x$ for large $x$, so the total execution and all IDs and record addresses have a fixed polynomial range in $x$. The top level has
$\ell_{\rm main}t\le\operatorname{bitlen}(x)+t=O(\log x)$.
Packet levels similarly satisfy $\ell s=O(\log Q^*+s)=O(\log x)$. Capacities, IDs and addresses therefore fit the fixed-headroom metadata words of \cref{sec:model}.
\end{proof}

%% file: sections/ownership.tex
\section{Owner-store representation for nested calls}\label{sec:ownership}

An ancestor callback must access its own per-vertex record while a descendant is active (\ref{e:realization}). A lower-generation call can update an enclosing engine before it returns. An ordinary stack of per-vertex records is therefore insufficient: the record at the head may belong to a descendant when a callback needs the ancestor's ceiling or discovery state. We attach an owner ID to every record and search for the requested owner.

The search costs only the number of live generations, because a vertex has at most one record per owner and store role. Main-recursion frames stay inside their engine's state, so their depth does not lengthen this search. Lazy creation and direct deletion links make the same representation support first touches and cleanup.

\begin{lemma}[Sparse owner-store simulation]\label[lemma]{lem:storage}
Consider an ideal engine execution on $N$ vertices using a fixed number of store roles. At most $L\le c(D+1)$ engine owners are live at any instant, including the finite base namespaces. Let $E$ count all logical store operations, including callback lookups, ceiling reads, first touches, epoch checks, and writes. Let $J$ count all namespace creations and $R$ all record creations over the execution. Suppose the ideal event account pays all remaining engine work.

The stores can be simulated deterministically in
\begin{equation}\label{eq:storage-time}
 O(N+J+LE+R)
\end{equation}
time and $O(N+L+R_{\rm live})$ live memory, besides the ideal payloads and their paid work. Access is correct even when an ancestor callback first touches a vertex while a descendant is live. Cleanup permits arbitrary owner order. If the ideal total bound is $AB^D$ times a nonnegative execution budget, the simulation changes that bound by at most a polynomial factor in $D$, absorbable in $A'(B')^D$.
\end{lemma}
\begin{proof}
For each vertex and store role, keep the head of a doubly linked list of live records. A record stores an owner ID, its payload, links in the vertex list, and links in its owner's touched-record list. There is at most one record for each owner, vertex, and role. Each owner has a descriptor and a touched-list head. Each generation descriptor allocates a fresh owner ID, which remains unique for the lifetime of its records.

A logical read or write names its owner explicitly. Scan the vertex-role list for that ID. Its length is at most $L$, so the search costs $O(L)$. If found, use exactly that payload. If absent, return the logical default, or create the record on a first touch according to the ideal store semantics. A created record is inserted at the vertex-list head and in its owner's touched list using $O(1)$ pointer operations. An ancestor callback performs exactly the same search and insertion even if descendant records are present. List position never determines ownership.

To clear an owner, traverse its touched list. Each record has direct links in the vertex list, so remove it in $O(1)$ time regardless of whether it is the head or whether younger records remain. Release its payload using the ideal cleanup semantics. This supports non-LIFO deletion. Each created record is inserted and removed at most once, contributing $O(R)$ total time. Default array heads cost $O(N)$ initialization for a fixed number of roles; owner descriptors cost $O(L)$ live space and $O(J)$ total creation work. Searches over all $E$ logical operations cost $O(LE)$. These facts prove \eqref{eq:storage-time} and the memory bound.

Correctness follows by induction on logical operations. For every owner and key, either the unique record contains the ideal value or its absence represents the ideal default. A lookup selects that owner's value; creation initializes it correctly; writes update it; and cleanup removes exactly the named owner's records. Other owners' values remain unchanged even when linked-list positions change. Synchronous callbacks are simply additional logical operations in this induction.

Every store operation, namespace creation, and record creation lies within the event budget, so the simulation multiplies total work by at most $L=O(D+1)$. For any fixed polynomial $p$, $p(D)B^D\le A'(B')^D$ for suitable absolute $A',B'>1$.
\end{proof}

In \cref{sec:bootstrap}, generation descent leaves at most one live owner per generation. The lemma therefore applies with $L=O(D+1)$ to the event count established there. A receiver pointer and an abort target in each generation descriptor route publications and stops to the intended enclosing call; the guarded construction determines their numerical meaning.

%% file: sections/calculus.tex
\section{Generation-uniform factor calculus}\label{sec:calculus}

\begin{remark}[Logarithms and integer evaluation]\label[remark]{rem:logarithms}
The factor identities use binary logarithms. The implementation evaluates constant-factor upper approximations by integer bit-length operations on scale metadata.
\end{remark}
 We use the definitions \eqref{eq:logs-intro}--\eqref{eq:phi-intro} for real $x\ge0$. Put $\psi(y)=\log_2(4+y)$. This increasing concave map has one positive fixed point $\rho\in(2,3)$. For sufficiently large absolute $x$, $L_1(x)\ge\rho$, and the sequence $L_i(x)$ decreases to $\rho$.

\begin{lemma}[Mass, convergence, and identities]\label[lemma]{lem:identities}
The exponents in each $f_d$ have total mass one. The product defining $\Phi$ converges to a finite positive value. For all $d\ge1$,
\begin{align}
 f_{d+1}(x)^2&=L_1(x)f_d(L_1(x)+2),\label{eq:shift}\\
 \frac{f_{d+1}(x)}{f_d(x)}
 &=\left(\frac{L_{d+3}(x)}{L_{d+2}(x)}\right)^{2^{-(d+2)}},\label{eq:adjacent}\\
 \Phi(x)^2&=L_1(x)\Phi(L_1(x)+2).\label{eq:phi-shift}
\end{align}
On the decreasing-iterate region,
\begin{equation}\label{eq:factorbounds}
 \sqrt{L_1(x)}\le\Phi(x)\le f_d(x)\le L_1(x).
\end{equation}
For each fixed $x$, $f_d(x)\to\Phi(x)$ as $d\to\infty$.
\end{lemma}
\begin{proof}
The exponent sum is $\sum_{i=1}^d2^{-i}+2\cdot2^{-(d+1)}=1$. At fixed $x$, all iterates lie in a compact interval bounded away from zero, so the weighted series $\sum_i2^{-i}\log L_i(x)$ converges absolutely. The shift $L_i(L_1(x)+2)=L_{i+1}(x)$ proves the first and third identities; cancellation proves the second.

The lower bound in \eqref{eq:factorbounds} retains only the first factor, since every other iterate is at least one. The upper bound uses exponent mass one and $L_i\le L_1$. The same bounds apply to $\Phi$. To compare $f_d$ with $\Phi$, cancellation gives the useful exact identity
\begin{equation}\label{eq:tailidentity}
 \log\frac{f_d(x)}{\Phi(x)}
 =2^{-(d+2)}\left[\log L_{d+2}(x)
  -\sum_{j\ge1}2^{-j}\log L_{d+2+j}(x)\right].
\end{equation}
This is nonnegative when the iterates decrease, and tends to zero at fixed $x$ by boundedness. This also proves the stated limit.
\end{proof}

Rounding a width or reserving quota slack changes its argument by a constant factor. The stability bound below is uniform over the generations.

\begin{lemma}[Uniform stability]\label[lemma]{lem:stability}
Fix absolute positive multipliers. The fixed-multiplier stability constants of all $f_d$ and of $\Phi$ can be chosen independently of $d$. Moreover, for one fixed $c\ge2$, if
\[
 q=\left\lfloor\frac{t}{c f_d(t)}\right\rfloor,
\]
then, for every sufficiently large absolute $t$ and every $d$,
\begin{equation}\label{eq:quota-stability}
 \tfrac12 L_i(t)\le L_i(q)\le L_i(t)\quad(i\ge1),
 \qquad \tfrac12 f_d(t)\le f_d(q)\le f_d(t).
\end{equation}
The lower factor $1/2$ can be replaced by another fixed positive constant when bounded dyadic rounding is included.
\end{lemma}
\begin{proof}
By \eqref{eq:factorbounds}, $f_d(t)\le L_1(t)$ uniformly. Thus taking the quota loses only $O(\log L_1(t))+O(1)$ from its first logarithm, and $L_1(q)\ge L_1(t)/2$ above an absolute threshold. Concavity and $\psi(0)=2$ give
\[
 \psi(\theta y)\ge\theta\psi(y)\qquad(0\le\theta\le1).
\]
Monotonicity and this inequality propagate the factor $1/2$ through every later iterate. Multiplying with exponent mass one proves the factor statement. For a fixed multiplier in the argument, the first iterates are within one fixed ratio, and the same concavity argument propagates that ratio. Finite and infinite exponent mass one then give stability of $f_d$ and $\Phi$. All low-scale exceptions can be bounded on one fixed compact interval.
\end{proof}

An acquisition uses a lower call whose level depends on the original quota. The packet estimate bounds that entire call, including its level term, by the desired acquisition rate.

\begin{lemma}[Uniform packet bound]\label[lemma]{lem:packet}
Let $F=f_d$, $g=f_{d-1}$, and let $s$ be within fixed positive multipliers of $F(q)$, with $s$ above the absolute base threshold. Then
\begin{equation}\label{eq:packet}
 s+\left(1+\frac{L_1(q)}s\right)g(s)+(\log_2s)^2=O(F(q))
\end{equation}
with an absolute constant independent of $d$.
\end{lemma}
\begin{proof}
The shift identity and monotonicity imply
\[
 g(F(q))\le g(L_1(q)+2)=\frac{F(q)^2}{L_1(q)}.
\]
Uniform fixed-multiplier stability gives the same inequality for $g(s)$ up to an absolute factor. Consequently $(L_1(q)/s)g(s)=O(F(q))$. Also $g(s)\le L_1(s)\le s$ at the legal scales, and $(\log_2s)^2\le s$ above a fixed threshold. Since $s=\Theta(F(q))$, their sum is $O(F(q))$.
\end{proof}

At the top of a generation, the analogous balance chooses a width from the measured graph size. This converts the local raw bound into the factor $f_d(x)$.

\begin{lemma}[Uniform root balance]\label[lemma]{lem:balance}
If $t=\Theta(f_d(x))$ with fixed multipliers, then, above an absolute scale,
\begin{equation}\label{eq:balance}
 t^2=\Theta\bigl(L_1(x)f_{d-1}(t)\bigr),\qquad
 t+\frac{L_1(x)}t f_{d-1}(t)+(\log_2t)^2=O(f_d(x)),
\end{equation}
uniformly in $d$.
\end{lemma}
\begin{proof}
Put $y=L_1(x)+2$. From \eqref{eq:factorbounds},
$\sqrt{L_1(x)}\le f_d(x)\le L_1(x)$. Hence $L_1(t)$ and $L_1(y)$ are within absolute positive ratios, independently of $d$. Propagating these ratios with the concavity argument from \cref{lem:stability} and using exponent mass one gives
$f_{d-1}(t)=\Theta(f_{d-1}(y))$. Apply the finite-generation shift identity \eqref{eq:shift} with index $d-1$. The squared-log term is $O(t)$ at the legal widths, proving the second assertion.
\end{proof}

\subsection{An explicit integer factor evaluator}

We compute a constant-factor upper bound on $f_d$ using the metadata operations of \cref{sec:model}. For an integer scale $x\ge0$, set
\[
 a_1=\lceil\log_2(x+2)\rceil,\qquad
 a_{i+1}=\lceil\log_2(4+a_i)\rceil,\qquad
 b_i=\lceil\log_2a_i\rceil.
\]
All these are exact integer bit-length operations:
$a_1=\operatorname{bitlen}(x+1)$,
$a_{i+1}=\operatorname{bitlen}(a_i+3)$, and
$b_i=\operatorname{bitlen}(a_i-1)$, with $\operatorname{bitlen}(0)=0$.
For requested generation $d$, form
\begin{equation}\label{eq:integer-factor}
 Q=\sum_{i=1}^{d}b_i2^{d+1-i}+b_{d+1}+b_{d+2},\quad
 p=\left\lceil Q/2^{d+1}\right\rceil,\quad
 \widehat f_d(x)=2^p.
\end{equation}

\begin{lemma}[Integer evaluator]\label[lemma]{lem:evaluator}
The routine \eqref{eq:integer-factor} satisfies
\[
 f_d(x)\le\widehat f_d(x)\le8f_d(x).
\]
At $d=O(\ls(x+2))$ it uses $O(d+1)$ word operations, assuming bit length as a metadata primitive, and $O(\log(x+2))$-bit words with fixed headroom.
\end{lemma}
\begin{proof}
We first show $L_i(x)\le a_i\le2L_i(x)$. The base holds since $L_1(x)\ge1$. If it holds at $i$, monotonicity gives the lower bound at $i+1$, while
\[
 a_{i+1}\le\log_2(4+2L_i)+1
          \le L_{i+1}+2\le2L_{i+1},
\]
since $L_{i+1}\ge2$. Hence $L_i\le2^{b_i}\le4L_i$. The exponents sum to one, so the unrounded dyadic product lies between $f_d$ and $4f_d$. The final ceiling of its binary logarithm costs at most another factor two.

The numerator $Q$ is accumulated by integer shifts and additions. Its bit length is $O(d+\log\log(x+2)+1)$; the denominator uses $d+2$ bits. These fit the stated word budget at the selected depth. The output is at most $8L_1(x)$ on the asymptotic region. Every loop iteration uses a constant number of bit-length, shift, and addition operations. Fixed-size low-scale exceptions can be handled by the finite base.
\end{proof}

For a generation-$d+1$ acquisition, choose the integer quota
\[
 \BeginAccSupp{method=escape,ActualText={q = floor(t / (4096 * fhat_d(t))).}}%
 q=\left\lfloor\frac{t}{4096\widehat f_d(t)}\right\rfloor.
 \EndAccSupp{}%
\]
The lower packet width is a fixed upward dyadic multiple of $\widehat f_d(q)$. The margin $4096$ absorbs the evaluator's factor eight and rounding. The recurrence $T_{d+1}=2^{64(T_d+1)^2}$ remains sufficient after choosing $T_4$ large enough: since $\widehat f_d(t)\le8L_1(t)$, for all such legal widths
$L_1(q)\ge\tfrac12\log_2t$. Thus
\[
 \widehat f_d(q)\ge f_d(q)\ge\sqrt{L_1(q)}
 \ge\sqrt{32}\,(T_d+1)>T_d.
\]
The quota is large enough for the fixed local conventions, and $3q+1\le t$. Likewise $x\ge X_d$ implies a main width chosen at or above $\widehat f_d(x)$ is at least $T_d$. Uniform stability absorbs the fixed-factor rounding, while the acquisition budget in \cref{sec:bootstrap} bounds parameter evaluation.

%% file: sections/selector.tex
\section{Integer depth selection}\label{sec:selector}
\subsection{Exact selection without constructing a tower}

Guarded integer arithmetic evaluates \eqref{eq:depth} within the measured scale: each threshold is constructed after verifying that it fits below $x$.

\begin{algorithm}[H]
\caption{\textsc{MaxLegalDepth}$(x)$}\label{alg:selector}
\begin{algorithmic}[1]
\If{$x<T_4$} \State \Return \textsc{Small} \EndIf
\State $L\gets\lfloor\log_2x\rfloor$, $u\gets T_4$, $d\gets3$
\While{$u<x$}
  \State $a\gets u+1$
  \If{$a>\left\lfloor\lfloor L/64\rfloor/a\right\rfloor$}
    \State \textbf{break}
  \EndIf
  \State $e\gets64a^2$, $u\gets2^e$, $d\gets d+1$
\EndWhile
\If{$d<4$} \State \Return \textsc{Small} \EndIf
\State \Return $d$
\end{algorithmic}
\end{algorithm}

\begin{lemma}[Selector correctness and word bound]\label[lemma]{lem:selector}
The procedure returns $D_*(x)$ when $x\ge X_4$, and returns \textnormal{\textsc{Small}} otherwise. It performs $O(D_*(x)+1)$ threshold tests on the asymptotic region, uses $O(1)$ live scalar metadata, and materializes no integer larger than $x$ except fixed code constants.
\end{lemma}
\begin{proof}
For positive integer $a$, the guard passes exactly when $64a^2\le L$. This equivalence follows by integer division, without forming the square. Only after it passes do we form $e$; then $e\le L$ and $2^e\le x$. The preliminary test $u<x$ also ensures $a=u+1\le x$. Each value fits in $O(\log(x+2))$ bits. A shift representing $2^e$ is performed only when its highest set bit fits.

Initially $u=T_4$ and $d=3$. At any subsequent loop entry the invariant is $u=T_{d+1}$. A successful step replaces $u$ by $T_{d+2}$ and increments $d$, preserving the invariant. Since
\[
 2^{64(u+1)^2}\le x\quad\Longleftrightarrow\quad
 64(u+1)^2\le\lfloor\log_2x\rfloor,
\]
the first failed guard rejects exactly the next illegal threshold. All later thresholds are larger. The extra stop $u\ge x$ is safe because the next threshold is strictly greater than $u$. Thus the returned generation is the largest legal one; no successful step gives precisely the small-core case. The number of successes is $D_*-3$, followed by one failed test. Bit length is computed once; even an $O(\log(x+2))$ implementation of that initial operation is covered by the additive linear core cost.
\end{proof}

\subsection{Truncation at maximal depth}

The upper threshold bounds the iterates beyond depth $d$. Repeated logarithms reduce that threshold by one generation at a time.

\begin{lemma}[Stripping one threshold]\label[lemma]{lem:stripping}
Let $R(y)=256(y+1)^2$ and $A_0=R(T_4)$. For $j\ge5$,
\begin{equation}\label{eq:strip}
 \psi(R(T_j))\le R(T_{j-1}).
\end{equation}
If $d=D_*(x)$, then $L_{d-2}(x)\le A_0$, and all later iterates are at most $A_0$.
\end{lemma}
\begin{proof}
For $y\ge1$, $4+256(y+1)^2\le2048y^2$. Hence
\begin{align*}
 \psi(R(T_j))
 &\le11+2\log_2T_j\\
 &=11+128(T_{j-1}+1)^2\\
 &\le256(T_{j-1}+1)^2=R(T_{j-1}).
\end{align*}
The last inequality holds since $128(T_{j-1}+1)^2\ge11$.

Maximality gives $T_{d+1}\le x<T_{d+2}$. Therefore
\[
 L_1(x)<\log_2(2+T_{d+2})\le R(T_{d+1}).
\]
Apply \eqref{eq:strip} $d-3$ times. The index of $L$ advances from $1$ to $d-2$, while the threshold index falls from $d+1$ to $4$, proving $L_{d-2}(x)\le A_0$. Since $A_0>\rho$ and $\psi(A_0)\le A_0$, monotonicity keeps every later iterate below $A_0$. When $d=4$, this argument uses one application, so the boundary generation is included.
\end{proof}

\begin{proof}[Proof of \cref{thm:truncation}]
At these scales the iterates decrease, so the lower bound follows from \cref{lem:identities}. The first $d$ factors of $f_d$ and $\Phi$ agree. Every factor in the infinite tail is at least one, so
\[
 \frac{f_d(x)}{\Phi(x)}
 \le\bigl(L_{d+1}(x)L_{d+2}(x)\bigr)^{2^{-(d+1)}}
 \le A_0^{2^{-d}}
\]
by \cref{lem:stripping}. Finally, $D_*(x)\to\infty$ because every fixed threshold is eventually below $x$.
\end{proof}

\subsection{Depth and generation overhead}

\begin{proof}[Proof of \cref{lem:height}]
The lower recurrence bound $T_{j+1}\ge2^{T_j}$ implies
\[
 \ls T_{4+r}\ge r+\ls T_4.
\]
For the upper bound choose an absolute
$A\ge\max\{2^{20},4096(T_4+1)^2\}$, put $E_0=A$, and set $E_{r+1}=2^{E_r}$. We claim
\begin{equation}\label{eq:towerbracket}
 T_{4+r}\le\frac{\sqrt{E_r}}{32}.
\end{equation}
The choice of $A$ proves the base. If $T\le\sqrt E/32$, then $T+1\le\sqrt E/16$ and
\[
 2^{64(T+1)^2}\le2^{E/4}\le\frac{2^{E/2}}{32}
                         =\frac{\sqrt{2^E}}{32},
\]
where the middle inequality holds for our large $E$. This proves the induction. Consequently $\ls T_{4+r}\le r+\ls A$. Since $X_d=T_{d+1}$, we have
\[
 d-3+\ls T_4\le\ls X_d\le d-3+\ls A.
\]
For $X_d\le x<X_{d+1}$, apply monotonicity of log-star at both endpoints to conclude \eqref{eq:height}.
\end{proof}

Hence $B^{D_*(x)}=\exp(O(\ls x))$.

%% file: sections/consequences.tex
\section{Logarithmic and worst-case consequences}\label{sec:consequences}

\begin{remark}[Bounded numerical complexity]\label[remark]{rem:parameters}
For bounded $K$, the active core has bounded size. Its exact solution and full-label recovery therefore take $O(m+n)$ total time.
\end{remark}

The first two logarithms determine the leading exponents. All later iterates, together with the generation coefficient, contribute only to the $o(1)$ on the second logarithm.

\begin{corollary}[Leading logarithmic factors]\label[corollary]{cor:simple}
The factor in \eqref{eq:main} is
\begin{equation}\label{eq:simple}
 \Phi(K)\exp(O(\ls(K+2)))
   =\sqrt{\log_2 K}\;\cdot\;(\log_2\log_2 K)^{1/4+o(1)}
 \qquad(K\to\infty).
\end{equation}
Consequently the RAM bound can be written with $K$ times the factor on the right, using the convention in \cref{rem:parameters} at bounded $K$.
\end{corollary}
\begin{proof}
Separate the first two factors in $\Phi$:
\[
 \log\Phi(x)=\tfrac12\log L_1(x)+\tfrac14\log L_2(x)
                  +\sum_{i\ge3}2^{-i}\log L_i(x).
\]
The tail is nonnegative and at most $\tfrac14\log L_3(x)$ because the iterates decrease. This is $o(\log L_2(x))$. Further, $\ls(x+2)=o(\log L_2(x))$: after a fixed number of logarithms, the remaining log-star count grows more slowly than the resulting unbounded logarithm. For example $\ls y=O(\log y)$ for large $y$, applied after three initial logs, suffices. Hence the exponential coefficient changes only the $o(1)$ in the exponent of $L_2$. Finally $L_1(x)\sim\log_2x$ and $L_2(x)\sim\log_2\log_2x$.
\end{proof}

The gain from choosing depth as a function of the input also exceeds the improvement of any fixed generation. The next corollary compares the full running-time factor, including the exponential coefficient, with each fixed $f_r$.

\begin{corollary}[Comparison with every fixed generation]\label[corollary]{cor:fixed}
For each fixed $r\ge5$,
\[
 \Phi(x)\exp(O(\ls x))=o(f_r(x)).
\]
Thus maximal legal depth improves asymptotically on every fixed generation, with the gain quantified relative to $\Phi$.
\end{corollary}
\begin{proof}
Apply \eqref{eq:tailidentity} with fixed $r$ and put $z=L_{r+2}(x)\to\infty$. Since every later iterate is at most $\psi(z)$,
\[
 \log\frac{f_r(x)}{\Phi(x)}
 \ge2^{-(r+2)}\log\frac{z}{\psi(z)}
 =\Theta_r(\log z).
\]
For fixed $r$, $\ls x=o(\log L_{r+2}(x))$: removing a fixed number of logarithms changes the iteration count by at most a fixed constant, and log-star grows slower than a logarithm. The loss $O(\ls x)$ in the exponential is therefore absorbed by the displayed divergence.
\end{proof}

For the worst-case consequence, apply the same solver directly to the normalized input.

\begin{corollary}[General and sparse worst-case bounds]\label[corollary]{cor:sparse}
The same guarded-prefix construction, applied after linear constant-degree normalization, gives a graph-input-uniform exact SSSP algorithm with bound
\[
 O\!\left((m+n)\Phi(m+n)
                  \exp(O(\ls(m+n+2)))\right).
\]
For $m=O(n)$ this is
$O(n\sqrt{\log n}(\log\log n)^{1/4+o(1)})$.
\end{corollary}
\begin{proof}
Apply the linear split directly to the reachable graph, without the active-core reduction. Start the reconstructed raw solver at the source with infinite enclosing ceilings and select its legal generation from $x=M+N+2=O(m+n)$. The root account, truncation estimate, and fixed-multiplier stability give the displayed bound, including the linear recovery of original labels. For sparse graphs use \cref{cor:simple}.
\end{proof}

This route applies the solver directly to the normalized graph and uses the size-based bound; \cref{thm:main} additionally uses the active-core lower bound to express the cost in terms of $K$.

%% file: sections/reuse.tex
\section{Connection to reusable seeded queries}\label{br:reuse-main}\label{sec:reuse}

\paragraph{Query model.}
Reusable seeded queries give a complementary application of the solver. The graph and its nonnegative weights remain fixed; each query supplies held seeds $\beta(v)\in[0,\infty]$ and asks for $\min_x\{\beta(x)+\dist_H(x,v)\}$ at each vertex.

The reusable structure separates calibration from query work (\cref{sd:thm:main,sd:thm:composition}). Retain one exact
auxiliary distance vector and a single-entry tree for the weighted graph $H$. Calibration provides a zero-reduced-cost path from each tree ancestor to every descendant. Thus a later improvement at a region's entry can be applied as a minimum, while its completed interior remains reusable. At a tree vertex $g$, the
seeded composition solves a child quotient $Q_g$ with $k_g$ vertices and
$m_g$ original arcs, where
\[
 \sum_g k_g=n-1,\qquad \sum_gm_g\le m,\qquad k_g\le h,
\]
and $h$ is the number of tree leaves. A fresh source with one arc per finite
seed turns a local seeded solve into ordinary SSSP.

The reduced weights and seeds are nonnegative semantic values held as pairs
$(A,B)$ representing $A-B$. Coordinatewise addition and comparison through
$A+D$ versus $C+B$ simulate the solver at constant numerical overhead.
Numerical equality is preserved exactly, so its hop-count and vertex-ID tie
rules remain valid. Final scalar answers are recovered by the seeded
construction's tight-path traversal. Lemma~\ref{br:pairs} states the precise
simulation, including its output boundary.

\paragraph{Using the guarded solver locally.}
One may use either the uniform solver or Fibonacci-heap Dijkstra for each
quotient. Before a uniform solve, parallel variable arcs can be reduced to
their held minimum after all original residual seed contributions have been
processed. The resulting graph has $O(k_g^2)$ size. Fixed-power stability of
$\Phi$ then gives local time
\[
 O\left(m_g+k_g+
 \min\{k_g\log(k_g+1),
       (\widehat m_g+k_g)F(k_g)\}\right),
 \quad F(x)=\Phi(x)\exp(O(\log_2^*(x+2))),
\]
where $\widehat m_g$ is the number of retained endpoint pairs. Interleaving elementary steps of the two solvers achieves the minimum up to a constant factor. The path-minimum backend adds
$O((m+n)\alpha(m+n+2)+h\log(h+1))$ time
(\cref{br:quotient}), where $\alpha$ is a one-parameter inverse-Ackermann bound. This shared cost determines the overall query bound; the local solver can be chosen separately. Workspace consists of the backend storage and the maximum workspace
of the selected quotient computation.

%% file: sections/guarded_base_appendix.tex
\section{The ordinary guarded base and four finite prefix transforms}
\label[appendix]{gb:appendix}

Ordinary Dijkstra growth inside the main recursion of Duan--Mao--Shu--Yin~\cite{duan2026faster}, followed by four prefix transforms, gives the generation-four base used in \cref{prop:base-realized}.

\subsection{The guarded engine and its base}\label{gb:interfaces}

\begin{contract}[Uniform guarded engine]\label[contract]{con:engine}
There are absolute constants $A,B>1$, a sufficiently large integer $T_4\ge16$, and one finite recursive deterministic routine $\GEN(d,\cdot)$ with the following properties. The legality thresholds $T_d$ and $X_d$ are defined in \eqref{eq:thresholds-intro}.
\begin{enumerate}[label=\textup{(E\arabic*)},leftmargin=*]
\item\label{e:raw} For every legal generation $d\ge4$, width $t\ge T_d$, and word-range-bounded level $\ell$ as in \cref{gb:inputs}, a normally returning raw call $X$ has exact guarded semantics and cost
\begin{equation}\label{eq:raw}
 \mathcal{T}_d(X;t,\ell)\le C_d\,u_X
       \bigl[t+\ell f_{d-1}(t)+(\log_2 t)^2\bigr].
\end{equation}
in the event convention of \cref{sec:parameters}, where $U_X$ is the exact returned output and $u_X=\max\{1,|U_X|\}$. Interrupted prefixes use the completion bound of \cref{sec:completion}.
\item\label{e:packet} A generation-$(d+1)$ WOS step uses a fixed-slack quota $q=\Theta(t/f_d(t))$ and a generation-$d$ packet of width $s=\Theta(f_d(q))$. If $t\ge T_{d+1}$, then $s\ge T_d$; the packet level is at most $1+L_1(q)/s$ up to fixed rounding constants. The call restarts from one root, publishes transitively as in \eqref{eq:union}, admits the incident-prefix completion, and respects native and foreign stops.
\item\label{e:coeff} The ideal-event raw coefficient satisfies a linear one-step recurrence:
\begin{equation}\label{eq:C}
 C_{d+1}\le B(C_d+1),\qquad C_d\le AB^d.
\end{equation}
\item\label{e:realization} At runtime depth $D$, one finite implementation keeps generation states separate and routes callbacks and aborts to their owners, including while descendants are live. The event count includes allocation, publication, logical lookup, parameter evaluation, and cleanup. \Cref{lem:storage} adds one $O(N)$ initialization and $O(D+1)$ per logical access. The resulting overhead and fixed polynomial metadata costs fit an absolute exponential bound $AB^D$.
\item\label{e:root} On a core of observable size $x\ge X_d$, the routine chooses a legal dyadic main width $t=\Theta(f_d(x))$ and solves the core in
\begin{equation}\label{eq:core-root}
 O(M+N)+AB^dM\left[t+\frac{L_1(x)}{t}f_{d-1}(t)
                              +(\log_2 t)^2\right]
\end{equation}
ideal events. Same-generation calls retain $t$ and change $\ell$; generation descent uses the fresh lower engine in \ref{e:packet}.
\end{enumerate}
\end{contract}

\begin{contract}[Fixed guarded base]\label[contract]{con:base}
\leavevmode
\begin{enumerate}[label=\textup{(B\arabic*)},leftmargin=*]
\item\label{b:keys} Labels and admissibility are as in \cref{gb:inputs}: keys order length, hop count, and vertex ID lexicographically, with a separate infinity tag. Edge extension uses one addition and a hop increment. Fresh roots have held finite offsets and frozen prefix ceilings. The main skeleton maintains SEAL and the ready frontier: keys below the bound outside the target are complete, and snapshots cover the required tight frontier.
\item\label{b:base} A fixed raw generation-four engine has exact guarded semantics, output thresholds $\Theta(t^3 2^{\ell t})$, and for every legal width and level used below,
\[
 \mathcal{T}_4(X;t,\ell)\le C_4u_X
          [t+\ell f_3(t)+(\log_2t)^2].
\]
A full root solves its local guarded problem. Costs follow \cref{sec:parameters}, with the base's sentinel convention for empty work.
\item\label{b:local} Incident lists, including reverse lists, are queried only after their vertex has been accepted and that acceptance published through every enclosing engine. Initialization, allocation, and packet control use only held entry data, queried data, and a fixed ambient ID/address universe. Entry ceilings are captured before publication can overwrite them. These rules hold recursively.
\item\label{b:stops} Publication follows the local label/predecessor/discovery update and precedes the next adjacency or queue action. Stops have explicit owners; foreign stops propagate after cleanup and cannot be interpreted as full failure. Exports copy held values before retirement. Cleanup is linear in created records/frames plus exported vertices.
\item\label{b:structure} Independently of acquisition speed, CAP/MERGE growth is a connected accepted arborescence within the target, meeting the old forest first at its sole collision vertex. Full FAILURE completes the guarded closure. Successful mass is $O(u_X)$, and connected edge-disjoint grouping leaves $O(u_X/q)$ groups. Outputs partition as in \cref{gb:cost}, where boundary arcs are charged on disjoint intervals for each tail. Source capacities hold, and distinct incoming origins satisfy
\begin{equation}\label{eq:Qbound}
 \sum_{Y\text{ in subtree}(X)}|Q_Y|
 \le u_X(\Delta+1+\ell_X/t).
\end{equation}
\item\label{b:physical} The representation in \cref{gb:cost} uses indexed BASE, within-engine frame stacks, rank selection, separated Merge, direct deletion, and a nonnegative block potential. Its seven cost accounts include engine/epoch creation, stores, captures, receiver links, and cleanup.
\end{enumerate}
\end{contract}

\subsection{Labels, entry conditions, and the exact return}
\label{gb:inputs}

Fix a directed adjacency-list multigraph with maximum in-degree and out-degree
$\Delta=O(1)$. Parallel arcs retain distinct IDs. A finite label at $v$ is
$(a,h,v)$, compared lexicographically, and
\[
 \operatorname{ext}_{(u,v)}(a,h,u)=(a+w_{uv},h+1,v).
\]
Infinity is a discrete greatest marker. Fix a root $r$, its held offset, and
ceilings $c[v]$ satisfying the root constraint. A walk is admissible when every
prefix label satisfies its endpoint ceiling. Write $\pi(v)$ for its minimum
admissible label, or infinity. Deleting a cycle decreases length or hops and
preserves all later upper ceilings. Replacing a prefix by a smaller admissible
prefix also preserves them. Thus finite minima are attained by simple walks,
every prefix of a minimum walk is minimum, and tight arcs
$\operatorname{ext}_{e}(\pi(u))=\pi(v)$ form a DAG. Write $u\leadsto v$ for
reachability in this DAG, including a path of zero edges.

All live finite labels $d[v]$ must be admissible walk upper bounds and only
decrease. A vertex is \emph{complete} when $d[v]=\pi(v)$. For a bound $b$ and
a materialized distinct nonempty source list $S$ with $d[S]<b$, set
\[
 T(b,S)=\{v:\pi(v)<b,\ s\leadsto v\text{ for some }s\in S\}.
\]
The entry condition is
\begin{equation*}
 \text{every incomplete }v\text{ with }\pi(v)<b
 \text{ has a complete }s\in S\text{ with }s\leadsto v.
 \tag{SEAL}\label{gb:seal}
\end{equation*}
A fresh single-root call satisfies this condition. An arbitrary seeded
frontier need not satisfy it. A frontier $\langle U,R\rangle$ for a target
means that every target vertex outside the complete set $U$ has a complete
tight ancestor in $R$; this includes target vertices already complete before
the call. For a queue record with immutable stored key $\kappa[v]$, readiness
means $\kappa[v]=\pi(v)$, which also forces $d[v]=\pi(v)$.

For width $t\ge16$ and level $\ell\ge0$, the source list must satisfy
$|S|\le t^2 2^{\ell t}$ at every raw-call entry. The output threshold is $t^3 2^{\ell t}$, and a nonbase
queue pulls at most $b_\ell=t2^{(\ell-1)t}$ records at once. Choose an integer
$k=\Theta(t/\log_2t)$ with $k\ge2$ and $3k+1\le t$; fixed small widths use
fixed legal constants. A normal return $(\beta,U,D)$ must satisfy
\begin{align}
 U&=\{v\in T(b,S):\pi(v)<\beta\},\quad d[U]=\pi(U),
       \label{gb:exact}\\
 &\langle U,\{v\in D:\kappa[v]=\pi(v)\}\rangle
       \text{ is a frontier for }T(b,S).\label{gb:ready}
\end{align}
The queue is disjoint from $U$, contains every residual original source and
required outgoing boundary head, and has $\beta\le d[v]\le\kappa[v]<b$.
For every admissible output-edge extension $z<b$, the final head label is at
most $z$ (the \emph{closure inequality}). A full return has $\beta=b$ and an
empty queue. A partial return has $\beta<b$ and
$|U|=\Theta(t^3 2^{\ell t})$.

There is one $O(M+N)$ graph/array setup. Metadata, including $\ell t$,
capacity bounds, frame IDs and actual counters, must fit $O(\log N)$-bit
words in the original ambient address universe. The arbitrary-level bound
applies throughout this word range, including calls with small reached mass.
Numerical operations are
held addition, copying and comparison. The local cost counts unit logical
owner-store accesses and unit requests to an immediate external receiver;
execution inside higher receivers is charged to those receivers. At fixed
finite depth the owner-store simulation has constant overhead. At runtime
depth $D$, its $O(D+1)$ lookup factor is applied once to the complete ideal
event count, as in \cref{lem:storage}.

\subsection{A finite main skeleton}
\label{gb:algorithm}

The operation $\operatorname{Relax}(u,e,b)$ constructs
$z=\operatorname{ext}_e(d[u])$ and succeeds exactly when
$z<b$, $z\le d[v]$, and $z\le c[v]$. On success it assigns the label and parent,
records first discovery, and invokes the immediate receiver before any
further queue or adjacency action. Equality is a successful relaxation; a
vertex counts toward a discovery quota only once.

Ordinary growth at an unowned source $x$ runs indexed Dijkstra from its held
current offset, stopping at $k$ distinct accepted vertices, first contact with
the successful forest, or exhaustion. Only locally accepted vertices enter
its heap. The incumbent tests implement an additional frozen entry ceiling:
initially a candidate must satisfy the entry incumbent; subsequently the
incumbent is exactly the last label published by this synchronous search.
CAP creates a new tree. MERGE adds its new vertices and arcs to the collided
tree using its existing handle. FAILURE puts $x$ in $Q$ and adds its accepted
vertices to a distinct list $W$. Already owned sources are skipped. Failed
roots are excluded from all propagated groups, even if a later search touches
them. At first collision only the collision vertex was previously owned.

Partition the successful forest by postorder traversal into connected,
edge-disjoint pieces of sizes between $k$ and $3k$, allowing shared attachment
vertices. Pending lists are spliced rather than copied. A final short
residual joins a neighboring piece. The construction takes time linear in the forest
size and duplicates only linearly many attachment records. Assign each
nonfailed source to its first incident piece and discard source-empty
pieces. This gives disjoint groups $P_j$ with connected edge-disjoint witness
trees $F_j$, $|P_j|\le3k$, and $p=O(|F|/k)$.

More explicitly, regard each successful component as an undirected tree
for this traversal. A vertex starts with its singleton pending piece and
splices child residuals one at a time, including their joining edges.
Immediately after a splice reaches $k$ vertices, emit that piece and reset
the pending piece to the attachment singleton. Before a splice both pending
pieces have fewer than $k$ vertices, so every emitted piece has fewer than
$2k$ vertices. The final residual has fewer than $k$ vertices; join it to
an emitted piece sharing an attachment, obtaining fewer than $3k$ vertices.
Each successful component has at least $k$ vertices and hence emits a piece.
The final join costs $O(k)$ once per component, and all other list splices
and attachment records cost linear total time. One source-assignment stamp
per vertex makes the first-incident-piece assignment linear as well.

Here is the remaining main control; all lists use first-encounter order and
explicit deduplication. Queue keys are snapshots, not pointers to live labels.
\begin{algorithm}[H]
\caption{\textsc{Raw}$(b,S,\ell;t,k)$}\label{alg:raw}
\begin{algorithmic}[1]
\State If $\ell=0$, run BASE described below. Otherwise form $(P_j,Q,W)$ by ordinary growth, initialize an empty block queue $D$, insert each group's live minimum pivot, set $\beta$ to their minimum or $b$, and set $U=\varnothing$.
\State While $D\ne\varnothing$ and $|U|\le t^3 2^{\ell t}$, pull records $I$ with separator $b_i$. Form $S_i$ from $I$ and, for every pulled designated pivot, all its group's members with live label $<b_i$; deduplicate.
\State Call $\textsc{Raw}(b_i,S_i,\ell-1;t,k)$, obtaining $(\beta_i,U_i,D_i)$. Delete every $U_i$ vertex from parent $D$, then consume $D_i$ by separated Merge, retaining the smaller snapshot on duplicates.
\State Remove $U_i$ vertices from groups, marking once every nonempty group whose designated pivot was removed. Scan all outgoing arcs of $U_i$ using $\operatorname{Relax}(\cdot,\cdot,b)$. Insert accepted heads with candidate $z\ge b_i$. An inserted group member smaller than an unmarked group's live pivot becomes its pivot.
\State For each marked group, scan its remaining list for a live minimum and insert that pivot. Set $\beta=\beta_i$, append $U_i$ to $U$, and repeat.
\State If $D$ is empty set $\beta=b$. Insert each original source with $\beta\le d[x]<b$. Form $W'=\{v\in W\setminus U:d[v]<\beta\}$, scan its outgoing arcs, and insert accepted heads with candidate $z\ge\beta$. Append $W'$ to $U$ and return $(\beta,U,D)$.
\end{algorithmic}
\end{algorithm}
BASE inserts $S$ into an indexed balanced search tree of individual snapshots.
While nonempty and output size is at most $t^3$, it extracts its minimum,
outputs that vertex, scans all outgoing arcs, and inserts/decreases accepted
heads. Its return bound is the remaining minimum \emph{after the last scan},
or $b$ if empty. Successor links permit sorted traversal of its return queue.
Normal return pops group/output memberships before exposing parent work;
returned queue memberships are consumed before parent duplicate lookup.

\subsection{Correctness, progress, and failed sources}
\label{gb:correctness}

\begin{lemma}[Target and source facts]\label[lemma]{gb:target}
Under \eqref{gb:seal}, every accepted speculative vertex lies in $T(b,S)$,
and $\min d[S]=\min\pi(S)$. After growth, every incomplete vertex below $b$
has a complete tight witness in $P=\bigcup_jP_j$. Every vertex of
$T(b,S)\setminus T(b,P)$ lies in $W$ and is complete.
\end{lemma}
\begin{proof}
For every incomplete $s\in S$, SEAL supplies a complete $z\in S$ with
$z\leadsto s$, so $d[z]=\pi(z)\le\pi(s)$; a complete $s$ uses itself.
Consequently $\min d[S]\le\min\pi(S)$, and the reverse inequality follows
from $d\ge\pi$. An incomplete live minimizer would moreover have a complete
$S$ witness with strictly smaller live label. A vertex outside
the target below $b$ is complete by SEAL. A valid relaxation into such a
vertex cannot strictly improve it. Equality would have a minimum prefix
through a target vertex, placing the head in the target too. Prefix
replacement proves this assertion even for a speculative incomplete tail.

Ordinary Dijkstra finalizes its local ceiling optimum at extraction: the
first unsettled prefix of a better admissible path would already have a
smaller heap key. An exhaustive search from a complete source completes
all its tight descendants below $b$, since their optimal prefixes satisfy
every entry-incumbent ceiling. Thus a complete witness of a still incomplete
vertex cannot have failed and belongs to $P$. Apply the same witness argument
to any target vertex; if it has no $P$ witness, its complete $S$ witness
failed and its exhaustive tree contains it in $W$.
\end{proof}

\begin{lemma}[Exact prefix and monotone progress]\label[lemma]{gb:prefix}
Every normal main call satisfies \eqref{gb:exact}--\eqref{gb:ready}, the closure
inequality, source and output bounds, and disjoint sibling output intervals.
Every child satisfies SEAL. Parent progress is nondecreasing, and all remaining
live group/queue labels and stored queue keys are at least its progress bound.
\end{lemma}
\begin{proof}
The induction tracks both live labels and the immutable keys stored in queue records. A stale record is safe when a ready tight-path witness covers it; this is the invariant that allows queue operations and label decreases to coexist. At a nonbase iteration let $R$ be the ready records and
$C(R)$ their tight descendants. For each remaining group define
$H_j=P_j\setminus(\operatorname{keys}(D)\cup C(R))$. Maintain:
\begin{enumerate}[label=\textup{(\roman*)},leftmargin=*]
\item $U$ is exactly the discharged complete prefix of $T(b,P)$;
$\langle U,R\cup P\rangle$ is its frontier, and $R\cup P$ also covers all
globally incomplete vertices below $b$ by complete tight witnesses;
\item live group/queue labels and snapshots are at least $\beta$, snapshots
are below $b$ and no smaller than live labels, and $U\cap D=\varnothing$;
\item each nonempty group has its designated pivot in $D$, with
$\kappa[p_j]\le d[v]$ for every $v\in H_j$;
\item every stale record, $\kappa[v]>d[v]$, is covered by a ready tight witness.
\end{enumerate}
The last clause is STALE. Initialization follows from the previous lemma
and live-minimum pivot insertion; initially no record is stale.

Pull leaves all unselected snapshots at least $b_i$. A ready witness below
$b_i$ is pulled. A complete group member already in $D$ is either pulled or,
if stale, covered by a ready witness using STALE. A member covered by $R$
reduces to the same case. Every other complete group witness is in $H_j$;
pivot dominance forces its pivot to be pulled and live augmentation includes
it. Thus child sources contain a complete witness for every incomplete
vertex below $b_i$. The same argument using the parent frontier covers every
remaining parent-target vertex below $b_i$, including already complete ones.
There are at most $(3k+1)b_\ell\le t^2 2^{(\ell-1)t}$ source proposals.
All their live labels are at least the previous value of $\beta$ and below $b_i$.
By the minimum identity in \cref{gb:target}, their minimum optimum key is at
least that value as well. Hence the inductive child prefix starts no lower than
the previous progress bound, and its returned $\beta_i$ cannot decrease that bound.

Child output is deleted before Merge. Child snapshots lie in
$[\beta_i,b_i)$, below remaining parent snapshots; retaining a smaller
duplicate cannot destroy readiness. Tight edges leaving completed child
output transfer its witnesses to ready boundary records. Inside $b_i$ the
child supplied those records; outside it the parent's scans supply them.
The child's closure inequality means a parent re-test below $b_i$ cannot
strictly lower a head. Every strict external decrease therefore immediately
gets its own inserted record. These facts preserve STALE: changed vertices
are either returned and deleted or covered by the child's ready frontier;
old ready coverage transfers across the newly discharged set.

Any remaining low queue/group vertex with live label $<\beta_i$ would be
in the child's exact target prefix: augmentation or ready coverage supplies
the required source witness. It would have been returned and removed.
This handles complete low vertices as well as incomplete ones. New external
records have value at least $b_i\ge\beta_i$. An unchanged necessary group
member retains pivot dominance; a changed one is removed, ready-covered,
or explicitly queued. Replacing an unremoved pivot by an inserted lower
live member only lowers its stored comparison bound. Removed pivots are
reselected by a live minimum; an unremoved pulled pivot returns as a residual
source record. This proves all iteration invariants.

At exit an empty queue forces every group to be empty, so $\beta=b$; otherwise
the output threshold stopped the loop. The exact discharged prefix is
$U=\{v\in T(b,P):\pi(v)<\beta\}$. A vertex in $W\setminus U$ with live key
below $\beta$ must be complete, since otherwise global witness coverage and
progress contradict its omission. The previous lemma then makes $U\cup W'$
exactly the original target prefix. Final residual-source insertion and
$W'$ boundary scans establish the required return frontier and closure.
Every output tail was scanned at its complete label under bound $b$ in a
child-output scan or final $W'$ scan. BASE establishes the same facts by
minimum extraction, with its bound read after scanning. Hop growth prevents
reinsertion of a settled vertex on a zero cycle. These give the induction
base and closure assertion used above.

Partial output exceeds $t^3 2^{\ell t}$. The last child and final $W$ add
only a constant multiple of that threshold, using
$|W|=O(k|S|)$. Thus its size has the stated order. Exact prefixes and
nondecreasing progress imply disjoint sibling outputs.
\end{proof}

\begin{lemma}[Incoming origins, including entry seeds]\label[lemma]{gb:origins}
In a same-engine main-call subtree rooted at $X$,
\begin{equation}\label{gb:qbound}
 \sum_Y|Q_Y|\le |U_X|(\Delta+1+\ell_X/t).
\end{equation}
\end{lemma}
\begin{proof}
An initial distinct source supplies one seed origin. Every later explicit
frontier origin is a boundary publication from a complete output tail;
group augmentation, Pull, residual-source insertion and Merge carry that
origin. Speculative decreases create no new origin. Lower fresh engines have their
own fixed optimum and are paid by their acquisition bounds, rather than
being included in this same-engine origin sum. For a fixed complete
tail, calls outputting it form an ancestor chain. Its publication ranges
$[\text{child input},\text{own input})$, together with its terminal BASE or
$W'$ range, are disjoint. Its candidate on a fixed primitive arc is fixed,
so that arc supplies at most one such publication. Other repeated tests are
still executed and charged below.

If $v$ fails in two full calls, those calls are ancestor-related, since both
output $v$. After failure in the ancestor $Y$, exclusion from all its groups
means a descendant source occurrence needs a new boundary insertion inside
$Y$. An earlier origin tail $u$ was output before entry to $Y$, with
$\pi(u)$ strictly below that earlier returned progress. Progress is
nondecreasing. All $Y$ sources have live key at least its entry progress;
SEAL and \cref{gb:target} make their minimum optimum at least that progress.
Consequently every vertex of its target has optimum at least that progress,
and $u\notin U_Y$. A descendant cannot publish again from that same tail.
The argument uses the completed origin tail and monotone progress, and applies whether or not the earlier candidate into $v$ was tight. Successive full failures inject into distinct incoming primitive
arcs, with one possible entry seed, giving at most
$(\Delta+1)|U_X|$ full occurrences. For partial calls,
$|Q_Y|\le |S_Y|<|U_Y|/t$. Outputs at a fixed depth are disjoint subsets of
$U_X$; summing at most $\ell_X$ nonbase depths proves \eqref{gb:qbound}.
\end{proof}

\subsection{Physical realization and the seven-account bound}
\label{gb:cost}

The algorithm uses the following storage representation. Arrays use lazy
epochs, touched lists, and held-value copies. Queues, groups, and accumulated
outputs have separate per-vertex membership heads. Within one engine,
membership records form frame stacks: ordinary executing-frame lookup
compares one head frame ID, without walking ancestors. Child group/output
memberships are popped before return. An output vertex is absent from its
returned queue, exposing the parent record for deletion. Merge pops each
child queue membership before parent lookup. Forest ownership and local
heap lifetimes end before recursive main calls. These facts justify the
stack discipline for ordinary frame work. Ancestor engine callbacks require
the owner-stores of \cref{lem:storage}, which locate the requested
engine's record during each cross-engine access.

A nonbase queue uses an AVL tree of ordered immutable-key intervals, each
holding an unsorted doubly linked block with direct item handles and block
pointers. Except for a sole small block, sizes lie in $[b_\ell/4,b_\ell]$.
Normalization collects a constant number of adjacent blocks and uses linear
rank selection to restore sizes in $[b_\ell/3,2b_\ell/3]$. Values retain their
original total order; distinct vertex IDs make queue keys distinct. For
block size $s$, use a sufficiently large constant times
\[
 \operatorname{dist}(s,[b_\ell/3,2b_\ell/3])
\]
as potential, with the lower-size exception for a sole small queue. A single
external insertion/deletion changes potential by $O(1)$. Underflow below
$b_\ell/4$ or overflow above $b_\ell$ releases $\Omega(b_\ell)$ credits;
its $O(b_\ell)$ reconstruction and AVL block changes are thereby paid.
Capacity is $A=O_\Delta(t^4 2^{\ell t})$, so
$\log(2+A/b_\ell)=O(t)=O(b_\ell)$. A sole small block is not rebuilt after
every deletion.

Insert, including duplicate inspection, costs $O(t)$ amortized. Deleting a
completed record through its handle costs $O(1)$ amortized; the normalization
potential pays for the associated block and AVL updates. Pull returns all records if size is at
most $b_\ell$, otherwise selects the first $b_\ell$ and their next-key
separator in $O(b_\ell)$ time. It reconstructs only a constant number of
boundary blocks. A separated child Merge has smaller keys than remaining
parent keys and child block parameter less than $b_\ell/3$. Pack consecutive
child blocks into $\Theta(b_\ell)$ batches, append the final short batch,
and delete duplicates through parent handles. The amortized cost is linear in the number of incoming
records. A BASE return is already sorted by successor links.
A small Merge can use normalization credits accumulated by earlier parent operations. All potentials are
nonnegative and zero at empty initialization.

The queue also stores neighboring-block links, its first
block, and its total record count. Interval endpoints are held separator
values, independent of the records that originally supplied them. Deleting
such a record therefore needs no search or recomputation of its separator.
The first interval extends to the bottom sentinel and the last to the call's
fixed upper bound. After Pull, extend the first remaining interval downward
to the bottom sentinel, so future admissible keys still have a destination.
Changing that extreme fence preserves the order of the other AVL nodes.
An Insert first inspects the vertex handle, keeps an existing no-larger
snapshot, or unlinks that occurrence and searches for the new interval.
An explicit Delete follows the handle directly; any resulting block repair
uses the normalization credits above.

A Pull on more than $b_\ell$ records collects initial blocks until it has
at least $b_\ell+1$ records. Block size is at most $b_\ell$, so this reads
$O(b_\ell)$ records. Linear rank selection retains the separator record and
returns exactly the $b_\ell$ smaller records. Its boundary reconstruction
and constant number of AVL changes cost $O(b_\ell)$, paid by the pulled
records; the reconstructed boundary blocks have zero potential.

If a separated Merge brings fewer than $b_\ell$ records, consume their child
memberships, delete parent duplicates through handles, and append the incoming
records to the first parent block, creating a sole block if necessary.
Their smaller keys already belong to that first interval, so this case makes
no per-record AVL search. Process their size changes as ordinary insertions
and deletions. A normalization triggered by these changes may spend credits
accumulated by earlier operations on the parent. For a larger Merge, ordered
child blocks of size less than $b_\ell/3$ form batches of
$\Theta(b_\ell)$ records; each new parent block's
$O(\log(2+A/b_\ell))=O(b_\ell)$ AVL insertion is paid by its batch.
Append the final short batch to its neighbor using the same small-Merge rule.
Internal movements during normalization receive no fresh insertion credits.
For the sole-small-block exception the potential is explicitly
$c\max\{0,s-2b_\ell/3\}$; all other blocks use the displayed distance
potential. Together these potentials pay for reconstruction across repeated
small Merges and deletions.

\begin{theorem}[Ordinary guarded raw bound $A_0$]\label[theorem]{gb:a0}
Under the inputs in \cref{gb:inputs}, the finite skeleton in
\cref{gb:algorithm} has the exact semantics of \cref{gb:prefix} and complete
local RAM cost
\begin{equation}\label{gb:rawbound}
 \mathcal{T}_X=O_\Delta\bigl(|U_X|[t+\ell_X\log_2t+(\log_2t)^2]\bigr).
\end{equation}
The constant is independent of ceilings, weights, numerical gaps, ambient
size, and any overestimate of reached mass by the legal level. Setup is
charged separately once.
\end{theorem}
\begin{proof}
Let $U$ be the parent output, with $u=|U|$. Let $U_i$ be the disjoint child outputs, $b_i$ their input bounds, $\beta$
the parent return bound, and $W'$ its final remainder. Define $N_I(Z)$ to be the set of
primitive outgoing arcs of $Z$ whose candidate from $\pi(\text{tail})$ lies
in $I$, including ceiling-rejected arcs. Put
$E_{\rm cut}=E(U)\setminus\bigcup_iE(U_i)$. The complete local accounts are:
\begin{center}\small
\begin{tabularx}{\textwidth}{@{}lYl@{}}\toprule
 & Events & Charge\\\midrule
1 & Ordinary growth, forest partition, initialization, initial pivots
  & $O(u\log t+t|Q|)$\\
2 & Group-source augmentation, excluding full replacement scans
  & $O(u)$\\
3 & Pull, Merge, deletion, memberships, output/edge scans, return, cleanup
  & $O_\Delta(u)$\\
4 & External interval Insert attempts
  & $O(t\sum_i|N_{[b_i,b)}(U_i)|)$\\
5 & Replacement pivots: partial / full parent
  & $O(u)$ / $O(u\log t+t|E_{\rm cut}|)$\\
6 & Original residual-source Insert attempts
  & $O(u)$\\
7 & Final remainder boundary Insert attempts
  & $O(t|N_{[\beta,b)}(W')|)$\\\bottomrule
\end{tabularx}
\end{center}
Successful growth is bounded by output size; failed growth and replacement scans use the failed-root and cut-edge terms.
Success/merge accepted mass is at most twice forest size, bounded both by
$|T|$ and by $O(k|S|)$. This is $O(u)$ for full calls, and for partial calls
$k|S|\le t|S|<u$. Failed mass is $O(k|Q|)$. Indexed growth heaps cost
$O(a\log(2+k))$ per accepted mass $a$, giving account 1, including
$O(u/k)$ initial pivot insertions. Distinct $W$ is deduplicated while its
already charged proposals are read; its final scan is $O(u)$.

A partial child pays every augmentation proposal by $k|S_i|\le|U_i|$.
A full child removes its pulled pivot: a group's last emptying scan occurs
once, and other full scans are paid with account 5. Pulled records number
$O(|U_i|)$ because a full child outputs its sources and a partial child
has $|S_i|<|U_i|/t$. Its returned queue contains only residual sources and
out-neighbors of output, hence $O_\Delta(|U_i|)$ physical records; it is empty
on a full return. This pays account 3, counting the same vertex again when
it moves through another ancestor. Explicit deletions number at most
$\sum_i|U_i|$ per parent.

For a full parent's group $j$, let $g_j$ count child iterations removing its
designated pivot and requiring replacement. Those removed vertices lie in
$g_j$ different child outputs on its connected witness tree. Deleting the
cut edges separates them, requiring at least $g_j-1$ edges. Witness pieces
are edge-disjoint and contained in $U$, so
$\sum_jg_j\le p+|E_{\rm cut}|$. Their edges need not remain tight.
For partial parents, each source is removed once and
$t|S|<u$. Original residual insertions occur only on partial returns.
This proves accounts 5 and 6. Accounts 4 and 7 charge all actual boundary
attempts, including duplicate tests, at their queue insertion cost.

For sufficiently large $C_\Delta$, induct on the explicit level with
$\mu_X=\min(\{b\}\cup\pi(S))$ and budget
\begin{equation}\label{gb:strong}
\begin{split}
 C_\Delta|U_X|(\ell_X+\log_2t)\log_2t
 +C_\Delta t\biggl(&|N_{[\mu_X,b)}(U_X)|+|E(U_X)|\\
                 &+\sum_{Y\text{ in subtree}(X)}|Q_Y|\biggr).
\end{split}
\end{equation}
BASE has $O(t^3)$ active records, so its ordered-tree operations cost
$O(u\log t)$ and fit. Child outputs and $W'$ partition $U$.
By progress, each child's interval starts at or above the previous parent
progress. Its lower boundary interval and the parent's $[b_i,b)$ interval
are disjoint subintervals of $[\mu_X,b)$. Child internal-edge sets together
with $E_{\rm cut}$ partition the internal-edge allowance. These partitions
pay accounts 4, 5's cut term, and 7. One spare level of $u\log t$ pays all
remaining direct terms except $t|Q_X|$, which is added to the subtree sum.
Applying \cref{gb:origins} and bounded degree to \eqref{gb:strong} gives \eqref{gb:rawbound}. Record allocation is proportional to the events above, and cleanup visits only touched records and their allocated arena pages.
\end{proof}

\subsection{Fresh-prefix completion and the finite A4 base}
\label{gb:finite}

This transform uses fresh lower engines. Given a growth root
$x$, quota $q$, and outer incumbents, lazily freeze
$c'[v]=\min(c[v],d_{\rm entry}[v])$ before publication, put only $x$ finite
at its held offset, and run the lower raw engine. The receiving main search
is suspended; this child is its only label writer. A vertex first touched
before its first publication therefore still exposes its entry incumbent.
Nested capture forces enclosing captures before the first outward write. Count the root once;
root initialization does not overwrite its enclosing predecessor with a
null parent. Every successful relaxation copies its label and nonroot
parent to the immediate receiver. That receiver records its own discovery,
invokes its outer receiver, and then tests its own collision/quota, so a
foreign ancestor stop takes precedence. Stop synchronously at first forest
contact or the $q$th distinct acceptance. Only a full normal root return
is FAILURE; internal partial returns are consumed normally.

If a prefix accepted $a\le q$ vertices, retain the full outgoing lists of
all queried tails, including the last partially scanned list, and all their
endpoints. There are at most $1+\Delta a$ vertices. Empty every other tail
list. Preserve IDs, root offset, ceilings, adjacency order, ambient word
universe and the original quota-derived parameter packet. Only accepted
finite tails are queried, initialization is lazy, and no choice uses the
unknown optimum or unqueried topology. Induction on actual operations
therefore matches the entire prefix, including rejected tests and later
reactivations. Arbitrary ceilings are essential: the old outer incumbent
registers need not themselves be paths in this small completion.

Disable actions external to the raw boundary only after matching the stopped
prefix, replacing its receiver by a bounded no-stop receiver. Native lower
stops remain enabled. The finite completion cannot discover an unlisted
vertex. Set $Q^*=C_\Delta(q+1)$, with slack exceeding its size. The original
packet has level $\max\{1,\lceil\log_2Q^*/s\rceil\}$, so its root threshold
exceeds the entire completion and its return must be full. Its actual full
cost bounds the actual prefix work. This comparison also pays an interrupted
amortized queue operation before its credits have been released. Abort
cleanup visits only created records/frames and exported vertices.

For $A_0$, take $s=\Theta(f_0(q))$, where
$f_0(q)=\sqrt{L_1(q)L_2(q)}$, enlarging fixed small packets as needed. Then
\[
 s+\frac{L_1(q)}s\log_2s+(\log_2s)^2=O(f_0(q)).
\]
\Cref{gb:a0} on the completion proves complete acquisition cost
$O_\Delta((1+a)f_0(q))$. Current parent labels strictly increase along parent
arcs, even after later decreases, so the exported arcs form an arborescence
at the exact callback point. A collision appends only fresh vertices and
one old endpoint, with $O(a)$ work. A full failure from a complete outer root
completes its needed tight suffixes: all satisfy the frozen upper ceilings.
Consequently the structural correctness and origin arguments above apply
with this acquisition replacing ordinary growth.

\begin{proposition}[Four fixed fresh-prefix transforms]\label[proposition]{gb:a4}
There is a fixed finite guarded generation-four program with exact raw
semantics and, for all its legal widths and word-range-bounded levels,
\[
 \mathcal{T}_{A_4}(X;t,\ell)=O_\Delta\bigl(|U_X|
                  [t+\ell f_3(t)+(\log_2t)^2]\bigr).
\]
Its constant is absolute at fixed $\Delta$. It uses the fresh-root
prefix construction above.
\end{proposition}
\begin{proof}
Write $r_0(t)=\log_2t$ and $r_j(t)=f_{j-1}(t)$ for $j\ge1$.
Write $u$ for the output size. Suppose $A_j$ has raw cost $O_\Delta(u[t+\ell r_j(t)+(\log t)^2])$.
Use its completion packet of width $s=\Theta(f_j(q))$. For $j=0$ use the
preceding calculation; for fixed $1\le j\le3$, the packet bound of
\cref{lem:packet} gives acquisition rate $O_\Delta(f_j(q))$.
Choose outer quota $q=\Theta(t/f_j(t))$ with fixed slack so $3q+1\le t$.
At sufficiently large legal widths,
$f_j(q)=\Theta(f_j(t))$ and $qf_j(q)=\Theta(t)$.

Successful/merge accepted mass is $O(u)$, while each failed attempt costs
$O(qf_j(q))=O(t)$. There are $O(u/q)$ groups. Thus account 1 becomes
$O(uf_j(t)+t|Q|)$, and the initial-group part of full replacement account 5
becomes $O(uf_j(t))$. Its cut-edge part and all other accounts stay unchanged.
Repeating \eqref{gb:strong} with level rate $f_j(t)$, followed by
\eqref{gb:qbound}, proves the raw $A_{j+1}$ bound. Applying this step for
$j=0,1,2,3$ gives $A_4$. Choose four fixed thresholds so every lower packet
is legal; a fixed routine handles the bounded-size global cases below them.
The five engine namespaces give constant overhead for callback and ceiling
accesses by \cref{lem:storage}. Root initialization, parameter evaluation,
and cleanup are bounded by the acquisition and main-call budgets above.
\end{proof}

This gives the base used in \cref{prop:base-realized}. The extension to variable depth is established in \cref{prop:reconstruct} using uniform packet estimates and the linear coefficient recurrence.

%% file: sections/bridge_proofs.tex

\section{Intrinsic cost and calibrated seeded composition}
\label[appendix]{br:section}

This appendix supplies the numerical lower bound used by the active core and the reductions used in the seeded-query extension. The active-core argument charges comparisons and additions on one execution, which is necessary because $\KCA$ optimizes their sum. The later results distinguish fixed-source reduction from reusable seeded computation and justify substituting the guarded solver into child quotients.

\begin{lemma}[Attainment of the numerical benchmark]\label[lemma]{br:attainment}
Fix a finite rooted topology. Among finite deterministic comparison--addition
programs correct for every nonnegative primitive weighting, the infimum
\[
 K=\KCA(G,s)=\inf_A\sup_w\bigl(C_A(w)+P_A(w)\bigr)
\]
is a finite nonnegative integer and is attained by one such program.
\end{lemma}
\begin{proof}
A finite unrolling of Bellman--Ford provides a finite upper bound, with
absence represented by a discrete tag. Every execution cost is a nonnegative
integer. Consequently a finite supremum of execution costs is itself an
integer. The set of finite worst-case costs of correct programs is thus a
nonempty subset of the nonnegative integers and has a least member. A
program whose cost is that member attains the infimum.
\end{proof}

\subsection{A same-execution active-edge charge}

To lower-bound the joint numerical cost, we choose weights for which many paths tie. Equality comparisons must distinguish perturbations of those paths, and additions must construct enough coefficient vectors on that same execution. A low-high order then ensures that sufficiently many active arcs participate. The structural input is the low-high-order theorem for
dominator trees: there is a preorder $\delta$ such that every $v\ne s$ either
has the arc $(\operatorname{idom}(v),v)$, or has incoming arcs $(a,v),(b,v)$
satisfying
\[
 a<_{\delta}v<_{\delta}b,
\]
where $b$ is not a descendant of $v$~\cite{georgiadis2012lowhigh}.

\begin{lemma}[Forward-DAG charge on one execution]\label[lemma]{br:forward-charge}
Let $\prec$ be a rooted order of a reachable directed multigraph $G$: the
source is first and every other vertex has an incoming arc from an earlier
vertex. Let $F$ contain all forward primitive arcs, with $m_F$ primitives,
$q_F$ endpoint classes, and $d_s$ source-out endpoint classes. For every
correct finite comparison--addition program $A$, there is one strictly
positive weighting $w$ on which
\[
 C_A(w)\ge m_F-n+1,\qquad P_A(w)\ge q_F-d_s.
\]
The two inequalities hold on the same execution, even for mixed sums,
arbitrary real literals, equality branches, and cross-output sharing.
\end{lemma}
\begin{proof}
Let $B$ be the vertex--arc incidence matrix of $F$, with column
$e_v-e_u$ for $u\to v$. Since $F$ is a reachable spanning DAG,
$\operatorname{rank}B=n-1$. Moreover,
\[
 \ker B=\operatorname{span}\{\chi_P-\chi_Q:
       P,Q\text{ are $s$--$v$ paths in $F$ for some $v$}\}.
\]
Indeed, fix an out-arborescence $T$ of $F$. For each non-tree arc
$e=(u,v)$, the vector $\chi_{T_u}+e_e-\chi_{T_v}$ is such a difference:
$T_u e$ is a path because $F$ is acyclic. These vectors are independent on
their respective non-tree coordinates, and their number is $m_F-n+1$.

Let $\mathbb F$ be the countable field generated by the finitely many real
literals of $A$. Choose potentials $\pi(s)=0$ and $\pi(v)$ algebraically
independent over $\mathbb F$, strictly increasing in $\prec$ and lying in
$(0,1)$. On forward arcs set $w_{uv}=\pi(v)-\pi(u)$. On every other arc
set $w_{uv}=\pi(v)-\pi(u)+\eta_{uv}$, where the $\eta_{uv}\in(2,3)$ are
jointly algebraically independent over $\mathbb F(\pi)$. All weights are
positive. Every forward path to $v$ has length $\pi(v)$, and any path using
an outside arc is longer, by telescoping.

Along the finite reached execution every register is affine in the primitive
weights, with nonnegative integer coefficients. Algebraic independence implies
that two equal registers have identical outside coefficients; their forward
coefficient difference lies in $\ker B$, and their constant terms agree.
A correct output for $v$ has zero outside coefficients, zero constant term,
and forward coefficient vector with incidence $e_v-e_s$.

Let $Z$ span the forward projections of the equality-comparison normals on
this execution. We claim $Z=\ker B$. Otherwise choose
$0\ne z\in\ker B\cap Z^\perp$. The path-difference identity gives an
endpoint $v$ whose forward path slopes $\chi_P\cdot z$ are not all equal.
For sufficiently small $\epsilon>0$, perturb only the forward weights by
$\pm\epsilon z$. Positivity, the strict outside-path gap, and every strict
comparison on the finite execution persist; each equality comparison remains
an equality. Thus both perturbations follow the same complete execution and
use the same affine output register. Yet the distance at $v$ has slope
$\min_P\chi_P\cdot z$ on the positive side and
$\max_P\chi_P\cdot z$ on the negative side. These differ, a contradiction.
Hence $Z=\ker B$, and each comparison supplies at most one normal, proving
$C_A(w)\ge m_F-n+1$.

To charge additions on the same execution, consider the coefficient vectors of all registers it creates. Project all
register coefficient vectors onto the forward coordinates and retain every
vector ever held. For a set $S$ define
\[
 D_B(S)=\operatorname{span}\{a-b:a,b\in S,\ Ba=Bb\}.
\]
Initially $S$ consists of zero and the forward unit vectors; outside inputs
and all literals project to zero. Thus $\dim D_B(S)=m_F-q_F$, contributed
exactly by parallel-class differences. Inserting one new vector in an already
occupied incidence fiber increases this dimension by at most one. The first
insertion into an unoccupied fiber increases it by zero. Each addition inserts
at most one vector. At termination the equality normals require dimension
at least $m_F-n+1$. Moreover, every vertex with no source-out class requires
an output in its distinct previously unoccupied fiber $e_v-e_s$, yielding at
least $n-1-d_s$ such first insertions. Therefore
\[
 P_A(w)\ge (m_F-n+1)-(m_F-q_F)+(n-1-d_s)=q_F-d_s.
\]
Discarding a physical register cannot invalidate this count, since the
argument retains its coefficient vector for analysis.
\end{proof}

\begin{proposition}[Active-core size charge]\label[proposition]{br:active-charge}
Delete every primitive $(u,v)$ whose head dominates its tail. In the remaining
graph let $m_s$ be the number of source primitives, $d_s$ the number of their
endpoint classes, $\sigma=m_s-d_s$, and $k$ the number of other primitives.
Then
\[
 \sigma+\frac{k}{2}\le K.
\]
\end{proposition}
\begin{proof}
An inactive arc occurs on no simple rooted path: reaching its tail has already
visited its head. Thus deletion preserves all distances and all simple rooted
paths. It also preserves the benchmark exactly: one simulation ignores deleted
inputs, and the reverse simulation supplies zero in their input locations.

Take a low-high preorder $\delta$ of the remaining graph's dominator tree,
and reverse every child list to obtain $\bar{\delta}$. Both orders are rooted.
The parent-arc alternative witnesses this in both orders. In the other
alternative the high predecessor $b$ is incomparable with $v$, so it becomes
a predecessor of $v$ after reversal. An active internal arc is forward in
both orders when its tail is an ancestor of its head, and in exactly one when
the endpoints are incomparable. The remaining ancestor direction would be
inactive. All source primitives are forward twice. Hence one order has
$m_F\ge m_s+k/2$. Also $q_F\ge n-1$ because the order is rooted.
For every competing program, Lemma~\ref{br:forward-charge} supplies one input
with
\[
 C_A+P_A\ge m_F-n+1+q_F-d_s
       \ge m_F-d_s\ge\sigma+k/2.
\]
Take its worst case and then minimize over programs.
\end{proof}

\subsection{Numerically ordered blocks and the addition bound}

The shared-hub example makes the role of shared additions explicit and will also separate the leaf parameter from $K$. Ordering spokes into blocks lets an inward scan stop early while limiting how many spokes require both an inward and an outward addition. The block containing the final hub minimizer determines that overlap.

\begin{lemma}[Final-minimizer block charging]\label[lemma]{br:rolling}
Consider the shared-hub graph with $k\ge1$ spokes and nonnegative weights
$s\to x:a$, $s\to v_i:b_i$, $v_i\to x:z_i$, and $x\to v_i:u_i$.
Let $1\le h\le k$, and partition the spokes into nonempty blocks
$B_1,\ldots,B_q$ of size at most $h$ satisfying the numerical invariant
\begin{equation}\label{br:eq:block-order}
 \max_{i\in B_p}b_i\le\min_{j\in B_{p+1}}b_j
 \qquad(1\le p<q).
\end{equation}
The order within each block is arbitrary, and equal values may occur in
different blocks. Initialize $H=a$. Scan the blocks in order, forming
$b_i+z_i$ and replacing $H$ by $\min\{H,b_i+z_i\}$ only when~$b_i<H$.
Finish the first block containing a failed strict test, then stop inward
scanning. With $H$ now final, output $b_i$ if $b_i\le H$; otherwise form
$H+u_i$ and output $\min\{b_i,H+u_i\}$.

Given the blocks, this algorithm computes all labeled distances using at
most $k+h-1$ additions, $O(k)$ comparisons, and $O(k)$ RAM time.
From arbitrary input values, blocks with $q=\lceil k/h\rceil$ can be
constructed using no numerical additions and $O(k\log q)$ comparisons
and RAM time, plus $O(k)$ initialization. Thus the complete algorithm uses
$O(k(1+\log\lceil k/h\rceil))$ comparisons and RAM time, while retaining
the same addition bound. Its workspace is $O(k)$.
\end{lemma}
\begin{proof}
First establish correctness from \eqref{br:eq:block-order}. Throughout
the inward scan, $H$ is the minimum of $a$ and the candidates formed so far.
At a skipped inward test $b_i\ge H_{\rm current}$, nonnegativity gives
$b_i+z_i\ge H_{\rm current}$. Such a candidate cannot improve either the
current or any subsequently lower hub value. At the first such test, every
spoke in a later block has value at least $b_i$ by
\eqref{br:eq:block-order} and is also safely skipped. The remaining spokes
in the current block are still tested, since their values may be smaller.
Therefore the final value is $H=\min\{a,\min_i(b_i+z_i)\}$. The spoke
distance is $\min\{b_i,H+u_i\}$; if $b_i\le H$, it is simply $b_i$.

If $H=a$, every spoke receiving an inward addition had $b_i<a$ and hence needs no outward
addition. Otherwise let $j$ be the inward candidate that first attains the
eventual final value $H<a$, and let $B$ be its block. A spoke in an earlier
block has $b_i\le b_j\le H$ and hence no outward addition. After block $B$,
the current hub already equals $H$, so every spoke receiving an inward
addition has $b_i<H$ and again receives no outward addition. Thus every
spoke receiving both additions lies in $B$. Spoke $j$ itself has $b_j\le H$
and receives no outward addition. There are at most $|B|-1\le h-1$ doubled
spokes and at most one first addition per spoke, proving the bound. A constant
number of comparisons per spoke suffices for this scan. The proof permits
ties and zero weights; $B$ is determined by the final minimizer and may
precede the stopping block.

To construct the blocks, use the total comparison key $(b_i,i)$, with the
spoke ID breaking numerical ties, and prescribe consecutive rank groups of
$h$ elements (the last may be smaller). Recursively split the list of target
groups into two halves. If the left half contains $r$ elements, deterministic
linear-time selection~\cite{blum1973selection} partitions the current records at rank $r$, after
which the two sides are processed recursively. Each level handles at most
$k$ records, and there are $\lceil\log_2 q\rceil$ levels. Selection and
partitioning therefore cost $O(k\log q)$ comparisons and RAM operations;
copying records and storing the blocks cost $O(k)$. For $q=1$, no selection
is needed. Only comparisons inspect the $b_i$ registers, so block
construction uses no numerical additions. A depth-first implementation
reuses partition storage and uses $O(k)$ workspace. The resulting blocks
satisfy \eqref{br:eq:block-order} without sorting their interiors.
\end{proof}

For the addition budget $k+r$, $0\le r\le k$, take
$h=\min\{r+1,k\}$. This gives at most $k+r$ additions and
$O(k(1+\log\lceil k/h\rceil))$ comparisons. More generally, a block
construction costing $Q(k,h)$ comparisons gives the transfer bound
$Q(k,h)+O(k)$; the scan's linear comparison bound is conditional on those
blocks being available. For $k=0$, the hub distance is the input $a$ and
the algorithm uses no numerical operations.

\subsection{Substituting a quotient solver}

Reduced costs are differences of held values, while the numerical model permits addition and comparison. Representing each difference by a pair preserves every comparison outcome with constant overhead. It allows the guarded solver to act on calibrated quotients.

\begin{lemma}[Signed-pair simulation]\label[lemma]{br:pairs}
An exact comparison--addition SSSP algorithm can be run on held pairs
$(A,B)$ representing nonnegative semantic weights $A-B$, with constant
overhead per numerical instruction. Its output is a pair for each distance.
The statement permits correlated inputs and a lexicographic label consisting
of numerical distance, hop count, and vertex ID.
\end{lemma}
\begin{proof}
Represent zero by $(0,0)$. Any fixed real literal $c$ is represented by
fixed nonnegative literals $(c_+,c_-)$ satisfying $c=c_+-c_-$; choosing
these literals is part of the finite code, not runtime subtraction. Add
coordinatewise, and compare $(A,B)$ to $(C,D)$
by a ternary comparison of $A+D$ with $C+B$. This uses two scalar additions
for a semantic addition, and two additions and one comparison for a semantic
comparison. Induction over execution shows that every semantic register and
comparison outcome agrees exactly with a run on the represented scalar
instance. Pairs contain materialized registers; temporary cross-sums can be
released immediately. A theorem valid on all independent nonnegative inputs
is valid on their correlated subfamilies as well. For lexicographic labels,
compare the pairs first, and on equality compare hop counts and then vertex
IDs, which remain ordinary integer metadata. Edge extension increments the
hop count as usual. Infinity remains a separate tag. The simulation returns held pairs. The seeded composition's final tight-path traversal converts its distance pairs to scalar outputs.
\end{proof}

\begin{proposition}[Quotient-solver composition]\label[proposition]{br:quotient}
Assume an exact comparison--addition SSSP solver $U$ with paid preprocessing,
cleanup, and RAM time
\[
 O\bigl((M+N)F(M+N)\bigr),\qquad
 F(x)=\Phi(x)\exp\bigl(c\log_2^*(x+2)\bigr)
\]
for an absolute $c\ge0$. Let a calibrated seeded composition have child
quotients $Q_g$, with $k_g$ vertices and $m_g$ primitive arcs. Write
$\widehat m_g\le k_g(k_g-1)$ for the number of distinct nonloop ordered
endpoint pairs. Then every quotient can be solved in
\[
 O\left(m_g+k_g+
 \min\left\{k_g\log(k_g+1),
          (\widehat m_g+k_g)F(k_g)\right\}\right)
\]
RAM time, with constant-overhead signed-pair arithmetic. The composition
therefore has per-query RAM bound
\begin{align*}
 O\Bigl(& (m+n)\alpha(m+n+2)+h\log(h+1)\\
 &+\sum_{g:k_g>0}
    \min\{k_g\log(k_g+1),(\widehat m_g+k_g)F(k_g)\}\Bigr).
\end{align*}
Using $U$ for every quotient gives the bound
\[
 O\bigl((m+n)\alpha(m+n+2)+h\log(h+1)+(m+n)F(h)\bigr).
\]
The preprocessing cost is additional. The workspace is the seeded backend's
workspace plus the maximum workspace of a chosen quotient solver.
\end{proposition}
\begin{proof}
First process every original residual seed contribution $R_e+c_e$, preserving
the distinct residuals of parallel arcs. For the
variable quotient contribution, retain only the minimum reduced weight for
each ordered endpoint pair, and discard nonnegative loops. Endpoint grouping
is topology-only and linear with integer vertex IDs, for example by stable
counting sorts; selecting the held minima uses $O(m_g)$ comparisons. Since
the calibrated reduced weights stay fixed, this reduction can also be retained
across queries, while residual seeds are processed anew in each query.

Add a fresh source with an arc to each finite quotient seed, using that held
seed as its weight. There are at most $k_g$ such arcs. All seed and reduced
edge values are semantically nonnegative. Reachability and absent outputs are
handled by the solver; the empty-seed case is immediate. The graph supplied
to $U$ has $M+N=O(\widehat m_g+k_g+1)=O(k_g^2+1)$.
Lemma~\ref{br:pairs} applies, including to zero cycles and ties.

For each fixed positive constant $a$, $F(a(k+1)^2)=O(F(k))$. To verify the product
part directly, $L_1(a(k+1)^2)\le c_a L_1(k)$ for a constant $c_a\ge1$.
The function $\psi(t)=\log_2(4+t)$ satisfies
$\psi(c_a t)\le c_a\psi(t)$, so induction bounds every iterate by the same
factor. Since $\sum_{i\ge1}2^{-i}=1$, the product ratio is at most $c_a$.
Also $\log_2^*(a(k+1)^2+2)=\log_2^*(k+2)+O(1)$. Thus the $U$ call costs
$O((\widehat m_g+k_g)F(k_g))$, in addition to the original-arc scan.
Fibonacci-heap Dijkstra costs $O(m_g+k_g\log(k_g+1))$ instead. Alternating
single paid steps of these two deterministic simulations and stopping when
the first finishes gives the minimum of their upper bounds with constant
overhead. Both return the same numerical distances; auxiliary tie choices
need not coincide. Their live storage~is the sum of the two simulations,
and disposing of the unfinished simulation costs at most the work already
spent allocating it. One may equivalently use a proved discrete cost selector.

Finally, $\sum_gm_g\le m$, $\sum_gk_g=n-1$, and $k_g\le h$. Summing the
local bounds and retaining the backend's independently charged
$h\log(h+1)$ term proves both assertions. The final original-graph tight-path
traversal materializes scalar outputs in $O(m+n)$ additional work. Sequential
quotient processing gives the stated workspace account. The resulting bound
is expressed in the quotient sizes and the leaf parameter $h$.
\end{proof}

%% file: sections/seeded/introduction.tex
\section{Reusable seeded composition}

Fixed-weight calibration lets the local shortest-path solver reuse numerical information across seed vectors.

Fix a directed graph $H=(V,E)$ with nonnegative real edge weights. A seeded shortest-path query supplies values $\beta(v)\in[0,\Inf]$ and asks for
\begin{equation}\label{sd:eq:query}
 d_\beta(v)=\min_{x\in V}\{\beta(x)+\dist_H(x,v)\}
 \qquad(v\in V).
\end{equation}
The value $\Inf$ marks an absent seed. Single-source and multisource shortest paths are special cases. More generally, an earlier computation may supply the entire seed vector, and later queries may depend on earlier answers. The graph and its primitive edge weights remain fixed.

A direct solution adds a temporary source with one outgoing arc per finite seed and runs a shortest-path algorithm. Repeating that procedure discards the opportunity to reuse numerical information about the fixed graph. At the other extreme, storing all pairwise distances uses quadratic space. We study an intermediate representation: one auxiliary distance vector and a rooted decomposition with one entry vertex per region.

The numerical issue is caused by cycles between regions. A child region may have been evaluated before a sibling discovers a better value for its entry vertex. Repeating the child computation is unnecessary if that entry value can be retained as a separate cap. Exact auxiliary distances make this possible: after reweighting, every ancestor in a suitable tree reaches each descendant at zero reduced cost. An improved ancestor value can therefore be propagated by a minimum, after the descendant's internal work has finished.

Each relaxation has two contributions: one from the mutable entry value of a child region and one from its permanent interior. The permanent contribution becomes a seed in the child quotient; the mutable contribution remains an arc. Solving this quotient resolves all cyclic interactions at that parent, after which the child entry values become permanent. A path-minimum structure supplies the interior contributions in any child-before-parent order, and the number of tree leaves controls the total quotient cost.

Fibonacci-heap Dijkstra on the local quotients gives the query bound below. \Cref{br:quotient} gives the optional substitution of the guarded solver.

\begin{theorem}[Reusable seeded queries]\label[theorem]{sd:thm:main}
Let $H$ be a finite directed multigraph with $n\ge1$ vertices, $m$ arcs and fixed nonnegative real weights. Choose a root $r$ reaching every vertex, and let its dominator tree have $h$ leaves. Let $T_{\rm dom}$ and $T_{\rm cal}$ be the actual RAM costs of computing that tree and the scalar distances $p(v)=\dist_H(r,v)$, respectively. There is a deterministic preprocessing and query algorithm with
\begingroup
\setlength{\abovedisplayskip}{0pt}
\setlength{\belowdisplayskip}{0pt}
\setlength{\abovedisplayshortskip}{0pt}
\setlength{\belowdisplayshortskip}{0pt}
\begin{equation}\label{sd:eq:prep}
 T_{\rm prep}=O(T_{\rm dom}+T_{\rm cal}+m+n),
\end{equation}
\begin{equation}\label{sd:eq:ram}
 T_{\rm query}=O((m+n)\alpha(m+n+2)+h\log_2(h+1)),
\end{equation}
\begin{equation}\label{sd:eq:ca}
 \CA_{\rm query}=O(m+n+h\log_2(h+1)).
\end{equation}
\endgroup
Every query returns all values in \eqref{sd:eq:query} as held scalar registers or infinity tags. Prepared storage and query workspace are both $O(m+n)$. These bounds include reading the seeds, initializing the query state, and writing all outputs. If calibration uses $P_{\rm cal}$ additions and comparisons, preprocessing uses $O(P_{\rm cal}+m+n)$ numerical operations in addition to the topology-only tree construction.
\end{theorem}

Here $\alpha$ is a standard one-parameter inverse-Ackermann upper bound for union by rank with path compression. The theorem permits arbitrary held seeds, including seeds correlated with the fixed weights or with previous answers. For $b$ queries, total RAM time is $T_{\rm prep}+bT_{\rm query}$, with $T_{\rm query}$ bounded by \eqref{sd:eq:ram}. Retaining all answer vectors requires an additional $O(bn)$ words.

For preprocessing, a linear-time dominator algorithm~\cite{alstrup1999} and Fibonacci-heap Dijkstra~\cite{fredman1987} give the explicit bound $T_{\rm prep}=O(m+n\log(n+1))$ with linear workspace. Array Dijkstra gives the simpler alternative $O(m+n^2)$. The dependence on $h$ is useful when the chosen dominator tree is narrow.

\paragraph{Related work.}
Tarjan's path-expression algorithms~\cite{tarjan1981} decompose directed graphs and combine local path computations; they yield almost-linear algorithms on reducible flow graphs. Dominator computation and tree-path evaluation have a long shared history~\cite{lengauer1979,alstrup1999}. We use the unary-chain and compressed-tree approach of Alstrup, Harel, Lauridsen, and Thorup, and prove the precise completion-order variant needed here. Fibonacci heaps~\cite{fredman1987} supply the local shortest-path bound. A-C trees~\cite{stefansson2026} organize shortest-path computation through recursively nested strongly connected components and measure complexity by nesting width. Our parameter is the number of leaves of a chosen auxiliary-root dominator tree, and our numerical preprocessing is reused across arbitrary seed vectors. The theorem expresses query complexity in terms of this leaf parameter; Section~\ref{sec:reuse} relates the resulting costs to the topology-specific comparison-addition benchmark~\cite{cai2026beyond}.

%% file: sections/seeded/calibration.tex
\section{Single-entry trees and numerical calibration}\label[appendix]{sd:sec:calibration}

We work in a comparison-addition RAM. A numerical instruction adds two held real values, compares two held real values, or copies one. Graph traversal, pointers, allocation and integer metadata are also paid in RAM time. An infinity tag is handled by control flow. Real values are never used as addresses, hashed, or inspected bit by bit.

Calibration is useful because the tree forces every route into a region through its entry. Exact calibration then guarantees a zero-reduced-cost path from the entry to every descendant.

For a rooted tree $D$, write $D_a$ for the subtree at $a$ and $a\preceq u$ when $a$ is an ancestor of $u$, including equality. For $a\preceq u$, let $D[a,u]$ be the set of vertices on the tree path from $a$ to $u$, and put $D(a,u]=D[a,u]\setminus\{a\}$. The tree itself need not be a subgraph of $H$.

\begin{definition}[Single-entry tree]\label[definition]{sd:def:tree}
A rooted tree $D$ on $V$, with root $r$, is single-entry for $H$ if every nonloop arc $(u,v)$ satisfies either $v\preceq u$, or $v\ne r$ and $\operatorname{parent}_D(v)\preceq u$.
\end{definition}

This property can be checked in $O(m+n)$ time with tree traversal intervals. It implies that the only vertex through which a path can enter $D_a$ from outside is $a$. Indeed, if an arc enters a proper descendant $v$ of $a$, both alternatives in \cref{sd:def:tree} force its tail to belong to $D_a$.

\begin{lemma}\label[lemma]{sd:lem:dom}
The dominator tree of a flow graph rooted at $r$ is single-entry.
\end{lemma}
\begin{proof}
For $v\ne r$, let $g$ be its immediate dominator. Every root-to-$u$ path followed by $(u,v)$ is a root-to-$v$ walk and must contain $g$. Since $g\ne v$, every root-to-$u$ path contains $g$. Thus $g$ dominates $u$, which means $g\preceq u$ in the dominator tree. Arcs to $r$ are returns to an ancestor. Loops satisfy the first alternative directly.
\end{proof}

The composition uses the single-entry property; the dominator tree provides a canonical tree satisfying it. Its construction, together with any choice of auxiliary root or tree, is part of preprocessing.

Compute and retain $p(v)=\dist_H(r,v)$. The assumptions make every $p(v)$ finite. Define, for analysis,
\begin{equation}\label{sd:eq:reduced}
 c(u,v)=p(u)+w(u,v)-p(v)\ge0,
 \qquad S=\max_{v\in V}p(v).
\end{equation}
An $O(m+n)$ certificate checks $p(r)=0$, all inequalities in \eqref{sd:eq:reduced}, and reachability from $r$ using tight arcs. Feasibility gives $p(v)\le\dist_H(r,v)$, while a tight path gives the opposite inequality. Thus the certificate verifies the entire calibration, including in the presence of zero cycles.

\begin{lemma}[Zero-cost descent]\label[lemma]{sd:lem:zero}
If $a\preceq u$, there is an actual $a$-to-$u$ path inside $D_a$ of reduced cost zero.
\end{lemma}
\begin{proof}
Take a shortest simple $r$-to-$u$ path; one exists because weights are nonnegative. The single-entry property forces it to visit $a$. Its suffix from $a$ to $u$ cannot leave $D_a$, since reentry would repeat $a$. Every prefix of the path is shortest, and every arc is tight for $p$. Telescoping \eqref{sd:eq:reduced} gives reduced cost zero for the suffix.
\end{proof}

For a query, initialize
\begin{equation}\label{sd:eq:q}
 q(v)=\beta(v)+S-p(v),
\end{equation}
with an absent tag if $\beta(v)=\Inf$. Every finite value is nonnegative. The answer in reduced coordinates is
\begin{equation}\label{sd:eq:zanswer}
 z_\beta(v)=d_\beta(v)+S-p(v).
\end{equation}
We will compute it as an ancestor minimum of final intrinsic labels $q$. The subtraction symbols in these displays describe values; \cref{sd:sec:pairs} gives the comparison-addition implementation.

%% file: sections/seeded/composition.tex
\section{Cyclic child composition}\label[appendix]{sd:sec:composition}

Let $g$ be an internal vertex of $D$. Its child quotient $Q_g$ has one vertex for each child of $g$. Write $c_e=c(u,v)$ for the reduced cost of an arc $e=(u,v)$. If an original arc $(u,v)$ goes from the subtree of one child $a$ to a different child subtree, the single-entry property forces its head $v$ to be that other child root. It contributes the quotient arc $a\to v$ with cost $c_e$. Parallel original arcs retain their identities. We call $g$ the owner of that arc. A return arc $(u,g)$ is owned by its head. Direct parent-to-child arcs and loops require no quotient arc.

If $k_g$ is the number of children and $m_g$ the number of quotient arcs, then
\begin{equation}\label{sd:eq:totals}
 \sum_g k_g=n-1,\qquad \sum_g m_g\le m.
\end{equation}
The quotients may be cyclic or irreducible.

\subsection{Mutable caps and permanent interiors}

Process the vertices of $D$ in child-before-parent phase order. At phase $g$, every proper descendant of each child $a$ already has a permanent intrinsic label. The child root $a$ itself remains a mutable cap awaiting the solve of $Q_g$. For an arc with tail $u\in D_a$ owned at $g$, define the held residual
\begin{equation}\label{sd:eq:residual}
 R_e=\min_{x\in D(a,u]}q(x).
\end{equation}
The path excludes $a$; its minimum is $\Inf$ if empty or entirely absent. Its entries are permanent before phase $g$.

For a cross-sibling arc $e=(u,v)$, the source value decomposes as
\begin{equation}\label{sd:eq:split}
 \min\{q(a),R_e\}+c_e
   =\min\{q(a)+c_e,R_e+c_e\}.
\end{equation}
Insert the constant $R_e+c_e$ into $v$'s quotient seed. Retain $a\to v$ as a quotient arc to handle the variable part. This separation is the reason a later change in $q(a)$ does not require reevaluating its interior.

Process a tree vertex $g$ as follows, finishing each child only after its label becomes immutable:
\begin{enumerate}[label=\textup{\arabic*.},leftmargin=*]
\item Start the seed of each child $v$ at its current $q(v)$. For each cross-sibling arc $(u,v)$, insert the constant $R_e+c_e$ into that seed.
\item Solve seeded shortest paths on $Q_g$ once, with its nonnegative reduced arc costs and the seeds from step 1. Replace every child cap $q(a)$ by the resulting quotient distance.
\item For each return $(u,g)$, with $u\in D_a$, update $q(g)$ by the candidate $\min\{q(a),R_e\}+c_e$.
\item Finish each child $a$ with its now immutable key $q(a)$. The residual queries in steps 1 and 3 precede these finishes.
\item At the overall root, finish $r$. Then propagate ancestor minima from root to leaves.
\end{enumerate}

The local solver computes actual shortest-path minima from its held seeds, rather than selecting an arbitrary solution of cyclic Bellman equations. Future ancestors affect an already processed region only through caps. The final traversal computes
\begin{equation}\label{sd:eq:finalcap}
 z(v)=\min_{x\in D[r,v]}q_{\rm final}(x).
\end{equation}

\subsection{Correctness}

The two inequalities needed for exactness have different sources. Walk witnesses show that the computation never underestimates a distance. The seed and edge inequalities then show that its final labels are no larger than any seeded path length.

\begin{lemma}[Walk witnesses]\label[lemma]{sd:lem:witness}
Every finite intrinsic label at a vertex $t$ has the form
\begin{equation}\label{sd:eq:witness}
 q(t)=\beta(x)+\operatorname{length}(W)+S-p(t)
\end{equation}
for an original seed $x$ and an actual original walk $W$ from $x$ to $t$. The same assertion holds for the final cap value $z(t)$.
\end{lemma}
\begin{proof}
Initially use the zero-length walk at $t$. A residual chooses a permanent label at an ancestor of $u$ and extends its witness to $u$ along the zero-reduced-cost path from \cref{sd:lem:zero}. A quotient relaxation appends a zero-cost descent from its source child root to the original tail, followed by the original arc. Thus every quotient path expands to an actual walk. A return update and a final ancestor cap have the same interpretation. Minima retain one existing witness. Concatenation may repeat vertices, which is allowed for walks and cannot create an invalid underestimate.
\end{proof}

\begin{lemma}[Final edge inequalities]\label[lemma]{sd:lem:feasible}
The final cap vector satisfies $z(v)\le\beta(v)+S-p(v)$ and $z(v)\le z(u)+c(u,v)$ for every original arc $(u,v)$.
\end{lemma}
\begin{proof}
Initial intrinsic labels remain candidates throughout, proving the seed inequalities. For a cross-sibling arc owned at $g$, let $a$ be the source child and set
\[
 A=\min_{x\in D[r,g]}q_{\rm final}(x),\qquad
 M=\min\{q_{\rm final}(a),R_e\}.
\]
The residual entries were already permanent at phase $g$, so $z(u)=\min\{A,M\}$. The quotient seed insertion and relaxation imply
\[
 q_{\rm final}(v)\le R_e+c_e,\qquad
 q_{\rm final}(v)\le q_{\rm final}(a)+c_e.
\]
The final ancestor pass also gives $z(v)\le A$. Hence, since $c_e\ge0$,
\[
 z(v)\le\min\{A,M+c_e\}
       \le\min\{A,M\}+c_e=z(u)+c_e.
\]
For a return $(u,g)$, the return update implies $q_{\rm final}(g)\le M+c_e$ and therefore $A\le M+c_e$. Also $A\le A+c_e$. These inequalities give $z(g)=A\le\min\{A,M\}+c_e$. For a direct parent-to-child arc, the ancestor pass gives $z(v)\le z(u)$ and nonnegative reduced cost suffices. A nonnegative loop also satisfies the inequality. \Cref{sd:def:tree} partitions the remaining arcs into precisely the cases above.
\end{proof}

\begin{theorem}[Exact cyclic composition]\label[theorem]{sd:thm:composition}
The phase algorithm computes $z(v)=z_\beta(v)$ for every vertex, for any single-entry tree, any exact auxiliary calibration, and any held nonnegative seed vector.
\end{theorem}
\begin{proof}
The walk witnesses imply $z(v)\ge d_\beta(v)+S-p(v)$. Conversely, follow the seed and edge inequalities of \cref{sd:lem:feasible} along any seeded path. They give $z(v)\le\beta(x)+\operatorname{length}(W)+S-p(v)$. Minimizing over paths gives the reverse inequality. Absent seeds and unreachable query vertices use the same tagged-minimum convention. The witness and inequality arguments apply equally to ties and zero-cost cycles.
\end{proof}

%% file: sections/seeded/backend.tex
\section{Path minima under delayed completion}\label[appendix]{sd:sec:backend}

Scanning a tree path for each residual in \eqref{sd:eq:residual} would repeatedly revisit the same vertices. We compress unary chains and maintain shared minima on the resulting tree. The structure needs only child-before-parent completion, so sibling regions can be processed in the order required by the quotient computation.

Each vertex is \emph{finished} once, after its children, when its immutable key is supplied. An operation $\Eval(u)$ on a finished vertex returns the minimum from $u$ to its highest finished ancestor, inclusive. An unfinished ancestor has no key in that component. The residual query therefore excludes the mutable cap of an unfinished ancestor.

\begin{theorem}[Bottom-up path minima]\label[theorem]{sd:thm:backend}
On a rooted tree with $n$ vertices and $h$ leaves, any child-before-parent sequence of finishes and $Q$ evaluations can be processed with
\[
 O(n+Q+h\log_2(h+1))
\]
key comparisons, $O((n+Q)\alpha(n+Q+2)+h\log_2(h+1))$ RAM work, and $O(n)$ space. Keys may be opaque held values accessed only through a comparison operation.
\end{theorem}

Chain compression and class minima, following Alstrup, Harel, Lauridsen, and Thorup~\cite{alstrup1997}, support arbitrary bottom-up finish orders. The proof below establishes this schedule and charges the operations of an ordinary union-find implementation.

\subsection{Unary chains}

Partition $D$ into singleton branching vertices and maximal chains of nonbranching vertices. There are $O(h)$ groups: at most $h-1$ vertices branch, and each chain starts at the root or a child of a branching vertex. The partition takes linear work and space.

Within one chain, every legal finish order is bottom-to-top, even if finishes on other chains intervene. Maintain a monotone stack of candidate minima. A new ancestor with key $k$ removes the stack suffix whose keys are strictly larger than $k$, and is pushed once. Each removed position is unioned with the new position; its disjoint-set representative stores the selected minimizing position. A query starting at a given chain position uses the representative of that position.

Inductively, an absorbed position is represented by the new, smaller ancestor; the surviving stack positions retain their previous minima. Thus a find returns exactly the minimum from the queried position through the finished part of its chain. Equal values may retain either witness. Each position is pushed once and popped at most once. There are $O(n)$ key comparisons and unions and $O(n+Q)$ finds over all chains. Union by rank and path compression give the stated inverse-Ackermann RAM cost~\cite{tarjan1975}; those finds do not compare numerical keys.

\subsection{The compressed tree}

When the bottom vertex of a chain finishes, activate its group and attach all completed child groups. The group's current key is the minimum over its finished chain segment. It can only decrease as ancestors on that chain finish. The group remains a component root until its parent group activates; by that time its top vertex has finished, so its key is immutable after linking. A branching singleton obeys the same rule with one finish.

Maintain classes of compressed vertices having the same minimizing ancestor. Each vertex points directly to its class; the class points to its canonical minimizing vertex. For each active component, a meldable max-heap contains the nonroot classes keyed by their canonical minima. The root class is excluded while its key is mutable.

When a parent of key $k$ receives a completed child component, insert the child's root class and meld its heap with the parent's. Repeatedly remove every maximal class with key greater than $k$ and absorb it into the parent's class. Remaining classes keep a minimum no larger than $k$. Decreasing the current root key performs the same absorption. By induction, the class pointer always names a true minimum on the path to the active component root.

Each compressed group creates a heap entry only when it links upward, at which point its key is immutable. Absorption removes that entry before merging the class. Thus there are $O(h)$ insertions, melds and deletions. Leftist heaps bound their total comparison and pointer work by $O(h\log(h+1))$. Root decreases cause at most one unsuccessful maximum test per finish, plus those deletions, contributing another $O(n)$ comparisons.

When classes merge, move the members of the smaller class into the larger one, reset their direct owner pointers, and set the merged class's canonical minimum to the parent. A moved member joins a class at least twice as large. The $O(h)$ compressed members therefore move $O(h\log(h+1))$ times in total. Class evaluation is constant time. At all times the groups, heaps, class lists and unary arrays contain $O(n)$ live records.

\subsection{Combining suffix and ancestor minima}

A query in a branching singleton uses its compressed class. A query in an unlinked unary group uses its unary suffix minimum. If the unary group has linked upward, compare that local suffix with the compressed minimum of the group's \emph{parent}. Using the group's full aggregate instead would include chain vertices below the query. The parent minimum includes precisely the remaining finished ancestors. This proves correctness and the operation bounds of \cref{sd:thm:backend}.

For the phase algorithm, keep every child $a$ unfinished throughout the residual queries at its parent. All proper descendants of $a$ have finished, so $\Eval(u)$ returns exactly the minimum on $(a,u]$. If $u=a$, return an absent tag without querying. Complete the quotient solve and all return updates before finishing the children. Every vertex finishes once and at most $m$ evaluations occur.

For example, with parent array $(\bot,0,0,1,1,2,2)$, the legal order $(3,4,5,6,1,2,0)$ finishes both sibling interiors before their roots. Such an order lies outside depth-first postorder, but satisfies exactly the ancestry condition used in the proof.

%% file: sections/seeded/realization.tex
\section{Addition-only realization and scalar output}\label[appendix]{sd:sec:pairs}

The reduced-cost formulas use subtraction only to describe values. Following \cref{br:pairs}, represent a semantic value $A-B$ by two held nonnegative scalar registers $(A,B)$. The needed operations are
\begin{align*}
 (A,B)+(C,D)&=(A+C,B+D),\\
 (A,B)\le(C,D)&\quad\Longleftrightarrow\quad A+D\le C+B.
\end{align*}
Thus addition uses two additions, and comparison uses two additions and one comparison. Copying a minimum copies a pair. Infinity remains a tag. The reduced arc cost is stored as $(p(u)+w(u,v),p(v))$ and an initial finite seed as $(\beta(v)+S,p(v))$.

Each pair is stored as two materialized scalars. Intermediate cross-sums are discarded after comparison, and a key update overwrites its two slots after computing the new values. The quotient solver maintains one active heap record per unsettled vertex and updates it by decrease-key. These rules keep the number of live pair records linear across repeated key changes.

To return scalar distances, we recover a tight seeded path for each finite value. This avoids evaluating a represented difference by subtraction. After the ancestor pass, adding the pair $(p(v),S)$ to the pair for $z(v)$ gives a pair for $d_\beta(v)$. Test, with pair comparisons, every finite seed equality $d_\beta(v)=\beta(v)$ and every tight finite arc equality
\[
 d_\beta(v)=d_\beta(u)+w(u,v).
\]
Start a graph search at the tight seeds. First discovery at a seed writes its held scalar $\beta(v)$; first discovery through a tight arc writes the held scalar sum $d_{\rm out}(u)+w(u,v)$. These writes give the correct distance by induction. Every finite output has a shortest seeded path whose initial seed and arcs are tight, so the search reaches it. Visited flags handle tight zero cycles. Absent outputs receive infinity tags. The pass uses $O(m+n)$ additions, comparisons and RAM operations and returns one held scalar for each finite answer.

%% file: sections/seeded/complexity.tex
\section{Preprocessing, quotient costs and space}\label[appendix]{sd:sec:cost}

Tree traversal intervals check the single-entry property. Classifying and routing arcs also takes linear time. For a cross-sibling arc $(u,v)$ its owner is $g=\operatorname{parent}_D(v)$; for a return its owner is $g=v$. Both owners are ancestors of $u$. During a second DFS keep the active ancestor at each depth. At $u$, the entry at depth $\operatorname{depth}(g)+1$ is the source child $a$. This gives $a$ in constant time per routed arc, without climbing its ancestor path. Store static quotient incidence and return lists once.

The calibration, its certificate, the shift $S$, reduced arc pairs and tree metadata can all be reused. A query initializes fresh labels and path-minimum state, performs the phases, and recovers the outputs.

Static routing and reduced-cost construction take $O(m+n)$ RAM time and numerical operations. Reading seeds and initializing fresh query state take $O(m+n)$ RAM time and $O(n)$ numerical operations.

The residual-minimum operations take $O((m+n)\alpha(m+n+2)+h\log(h+1))$ RAM time and $O(m+n+h\log(h+1))$ numerical operations. Writing $T_g$ and $P_g$ for the actual local quotient RAM and numerical costs, one solve per quotient contributes $\sum_g T_g$ and $\sum_g P_g$, respectively.

Returns, cap updates and scalar recovery take $O(m+n)$ RAM time and numerical operations. Cleanup takes $O(m+n)$ RAM time and $O(1)$ numerical operations.

Every tree edge contributes one child to exactly one quotient. The excess beyond unary quotients is concentrated at branching vertices; their total excess degree is $h-1$. This gives both aggregate bounds below.

\begin{lemma}[Leaf bounds for all quotients]\label[lemma]{sd:lem:leaves}
For a nonempty rooted tree with $h$ leaves,
\begin{align}
 \sum_g k_g^2&\le n-1+h(h-1),\label{sd:eq:squares}\\
 \sum_g k_g\log_2(k_g+1)&\le n-1+2(h-1)\log_2(h+1).\label{sd:eq:entropy}
\end{align}
\end{lemma}
\begin{proof}
Let $B$ be the branching vertices and put $x_g=k_g-1$ there. Counting tree edges gives $\sum_{g\in B}x_g=h-1$. Since unary vertices contribute zero to $k_g(k_g-1)$,
\[
 \sum_g k_g^2=n-1+\sum_{g\in B}(x_g^2+x_g)
 \le n-1+(h-1)^2+(h-1).
\]
This proves \eqref{sd:eq:squares}; a star attains equality. For \eqref{sd:eq:entropy}, unary vertices contribute at most $n-1$. Each branching vertex has $k_g\le h$, and
\[
 \sum_{g\in B}k_g=(h-1)+|B|\le2(h-1).
\]
Multiply by $\log_2(h+1)$. The same formulas hold for $h=1$, when the tree is a path, and for the one-vertex tree.
\end{proof}

\begin{proof}[Proof of \Cref{sd:thm:main}]
Use Dijkstra's algorithm with a Fibonacci heap on each quotient~\cite{fredman1987}. Insert its finite seeds, keep one handle per unsettled quotient vertex, and relax each outgoing arc when its tail is settled. Insert a previously absent vertex when first reached; later improvements use decrease-key. There are at most $k_g$ insertions and extract-min operations and at most $m_g$ decrease-key operations. Arbitrary nonnegative seeded values are equivalent to edges from a temporary source, so the usual correctness proof applies. Signed-pair addition and comparison give constant overhead per numerical primitive. Therefore
\[
 T_g,P_g=O(m_g+k_g\log(k_g+1)).
\]
Equations~\eqref{sd:eq:totals} and \eqref{sd:eq:entropy} bound their sum by $O(m+n+h\log(h+1))$. Combining this with \cref{sd:thm:backend,sd:thm:composition} and scalar recovery gives the RAM bound \eqref{sd:eq:ram} and the numerical bound \eqref{sd:eq:ca}.

All static arrays and incidence lists occupy $O(m+n)$ space. Each query uses $O(n)$ labels and backend records; a local quotient uses $O(k_g+m_g)$ space. Process quotients sequentially and release their temporary heaps after use. The pairs are materialized values and comparison temporaries are reclaimed, as described in \cref{sd:sec:pairs}. Thus the total workspace is $O(m+n)$, independently of the number of streamed queries. The workspace of the chosen tree and calibration algorithms is additional during preprocessing; the explicit choices stated after the theorem use linear space. Every query is correct for every held $\beta$, so adaptive choice of later queries changes neither the proof nor the worst-case bound.
\end{proof}

\begin{corollary}[Array implementation]\label[corollary]{sd:cor:array}
Replacing each local heap by an array scan gives per-query bounds
\[
 \RAM=O((m+n)\alpha(m+n+2)+h^2),\qquad
 \CA=O(m+n+h^2),
\]
with the same linear storage and workspace.
\end{corollary}
\begin{proof}
An array quotient solver costs $O(m_g+k_g^2)$. Sum \eqref{sd:eq:squares} and use $h\log_2(h+1)\le h^2$ for integers $h\ge1$.
\end{proof}

For $h=O(1)$, the query uses linear numerical work and almost-linear RAM time, regardless of dominator-tree depth.

%% file: sections/scope.tex
\section{Scope of reusable calibration}\label[appendix]{sec:scope}

For $b$ adaptive queries, total time is one-time preprocessing plus the sum of the $b$ query costs. Calibration
from the original source computes the original SSSP output, so its
weight-dependent cost belongs in any bound relative to $\KCA(G,s)$.
When the chosen dominator tree is a star, the root quotient can contain the original numerical difficulty. The leaf term is a structural cost: \cref{br:leaf-obstruction} gives active shared-hub graphs with $h=k+1$ and $K\le4k$, so no general $O(K)$ charge covers $h\log(h+1)$. The query bound therefore retains this term alongside the local solver costs.

The reduction rules also depend on the query model. Active-core deletion
preserves distances from its fixed root. For arbitrary seeds, the reusable
structure retains arcs that may return to that root: on $r\to v\to r$, the
return arc supplies the distance back to $r$ from a seed at $v$, and deleting
it changes that answer to infinity. Proposition~\ref{br:seed-obstruction}
formalizes this separation.

\begin{proposition}[Fixed-source reduction is not seed-universal]
\label[proposition]{br:seed-obstruction}
Deleting dominator-inactive arcs relative to a fixed root need not preserve
arbitrary seeded shortest paths. Adding a new source with a primitive arc to
every original vertex makes the new dominator tree a star.
\end{proposition}
\begin{proof}
Take two vertices $r,v$ and arcs $r\to v$ of weight $a\ge0$ and $v\to r$
of weight $b\ge0$. Relative to $r$, the return arc is inactive. With the
single seed $\beta(v)=0$ and $\beta(r)=\infty$, its deletion changes the
answer at $r$ from $b$ to $\infty$. Two vertices and two arcs suffice; a
one-vertex nonnegative loop cannot improve any seeded distance, and reaching
a second vertex while returning to the root requires both arcs.
For the second assertion, the direct arc from the new source $q$ to any
vertex $v$ avoids every other original vertex. Thus only $q$ and $v$
dominate $v$.
\end{proof}

\begin{proposition}[Separation of leaf count and intrinsic cost]
\label[proposition]{br:leaf-obstruction}
There is no absolute constant $c$ for which the dominator leaf count satisfies
$h\log(h+1)\le cK$ on all rooted graphs, even when restricted to source-class-normalized
active cores.
\end{proposition}
\begin{proof}
A source star with one primitive to each of $t$ leaves has $K=0$, since all
answers are already input registers, and $h=t$. This example disappears in
active-core reduction, so consider instead the shared-hub graph from
Lemma~\ref{br:rolling}. The following linear-cost algorithm needs no block
partition. Every nonsource vertex has a direct source arc;
its dominator tree is a star with $h=k+1$. The two directions between the hub
and each spoke are active, every nonsource vertex is incident to such an arc,
and source endpoint classes are singletons. Thus it already is a source-class-normalized
active core. Computing all $b_i+z_i$, selecting the hub minimum, and then
computing and testing all $H+u_i$ uses at most $4k$ additions and comparisons
for every nonnegative weighting. Hence $K\le4k$ while $h\log(h+1)$ is
$\Theta(k\log k)$, disproving the proposed inequality.
\end{proof}

\begin{proposition}[Preprocessing and adaptive queries]\label[proposition]{br:calibration-cost}
Let a fixed weighted graph admit the reusable seeded structure with
preprocessing cost $T_{\rm prep}$ and query bound $Q$. Any sequence of $b$
adaptive seed queries costs $O(T_{\rm prep}+bQ)$. If its calibration is
$p(v)=d_w(s,v)$, computing the calibration already computes the original
single-source output. Its weight-dependent computation therefore belongs to the total cost in any bound relative to $\KCA(G,s)$.
\end{proposition}
\begin{proof}
The structure is built once and a fresh query state is initialized for each
query. The query theorem holds for every held seed vector, including one
chosen from previous answers; summing its bounds proves the first assertion.
For the second, every required original distance is one of the calibration
registers. These registers depend on the primitive weighting and therefore
cannot be topology-only advice. If the weights change, the exact calibration
must be recomputed or separately maintained with its cost included.
\end{proof}